\documentclass[british,english,english,pre,onecolumn,superscriptaddress,notitlepage,nofootinbib]{revtex4-1}
\usepackage[T1]{fontenc}
\usepackage[latin9]{inputenc}
\usepackage{geometry}
\usepackage{color}
\usepackage{babel}
\usepackage{units}
\usepackage{textcomp}
\usepackage{amsmath}
\usepackage{amssymb}
\usepackage{graphicx}
\usepackage{esint}
\usepackage[unicode=true,pdfusetitle,
 bookmarks=true,bookmarksnumbered=false,bookmarksopen=false,
 breaklinks=false,pdfborder={0 0 0},backref=false,colorlinks=true]
 {hyperref}
\hypersetup{
 linkcolor=coolblack,citecolor=burntorange}

\makeatletter
\usepackage[T1]{fontenc}
\usepackage[latin9]{inputenc}
\usepackage{textcomp}
\usepackage{mathrsfs}
\usepackage{graphicx}
\usepackage{esint}
\usepackage{babel}
\usepackage{mathrsfs}
\usepackage{subfig}
\usepackage{booktabs}

\DeclareMathOperator*{\erfc}{erfc}

\definecolor{burntorange}{rgb}{0.8, 0.33, 0.0}
\definecolor{charcoal}{rgb}{0.21, 0.27, 0.31}
\definecolor{coolblack}{rgb}{0.0, 0.18, 0.39}

\makeatother

\begin{document}

\title{Learning may need only few bits of synaptic precision}

\author{Carlo Baldassi}

\author{Federica Gerace}

\author{Carlo Lucibello}

\author{Luca Saglietti}

\affiliation{Department of Applied Science and Technology, Politecnico di Torino,
Corso Duca degli Abruzzi 24, I-10129 Torino, Italy}

\affiliation{Human Genetics Foundation-Torino,  Via Nizza 52, I-10126 Torino,
Italy}

\author{Riccardo Zecchina}

\affiliation{Department of Applied Science and Technology, Politecnico di Torino,
Corso Duca degli Abruzzi 24, I-10129 Torino, Italy}

\affiliation{Human Genetics Foundation-Torino, Via Nizza 52, I-10126 Torino, Italy}

\affiliation{Collegio Carlo Alberto, Via Real Collegio 30, I-10024 Moncalieri,
Italy}
\begin{abstract}
Learning in neural networks poses peculiar challenges when using discretized
rather then continuous synaptic states. The choice of discrete synapses
is motivated by biological reasoning and experiments, and possibly
by hardware implementation considerations as well. In this paper we
extend a previous large deviations analysis which unveiled the existence
of peculiar dense regions in the space of synaptic states which accounts
for the possibility of learning efficiently in networks with binary
synapses. We extend the analysis to synapses with multiple states
and generally more plausible biological features. The results clearly
indicate that the overall qualitative picture is unchanged with respect
to the binary case, and very robust to variation of the details of
the model. We also provide quantitative results which suggest that
the advantages of increasing the synaptic precision (i.e.~the number
of internal synaptic states) rapidly vanish after the first few bits,
and therefore that, for practical applications, only few bits may
be needed for near-optimal performance, consistently with recent biological
findings. Finally, we demonstrate how the theoretical analysis can
be exploited to design efficient algorithmic search strategies.
\end{abstract}
\maketitle
\tableofcontents{}

\newpage{}

\section{Introduction}

It is generally believed that learning and memory in neural systems
take place via plastic changes to the connections between neurons
(synapses) in response to external stimuli, either by creating/destructing
the connections or by modifying their efficacy (also called synaptic
weight) \cite{hebb1949organization}. Although the details of how
these processes occur in neural tissues are largely unknown, due to
technical difficulties in experiments, this idea has inspired advances
in machine learning which in recent years have proven highly successful
in many complex tasks such as image or speech recognition, natural
language processing and many more, reaching performances comparable
to those of humans \cite{collobert2008unified,krizhevsky2012imagenet}.
The currently dominating paradigm in the machine learning field consists
of employing very large feed-forward multi-layer networks trained
with a large number of labeled examples through variants of the stochastic
gradient descent algorithm: the learning task is thus framed as an
optimization problem, in which the network implements a complex non-linear
function of the inputs parametrized by the synaptic weights, and the
sum over all the training set of the distance of the actual output
from the desired output is a cost function to be minimized by tuning
the parameters.

Despite the practical success of these techniques, it is rather unlikely
that actual brains employ the same gradient-based approach: real synapses
are generally very noisy, and the estimated precision with which they
can store information, although very difficult to assess conclusively,
ranges between $1$ and $5$ bits per synapse \cite{o2005graded,bartol2015}.
In other words, real synaptic efficacies might be better described
by discrete quantities rather then continuous ones. Conversely, the
gradient descent algorithm is well suited to solve continuous optimization
problems, and in fact machine learning techniques generally employ
synapses with at least $32$ bits of precision. Moreover, theoretical
arguments show that even in the simplest possible architecture, the
one-layer network also known as perceptron, the properties of the
training problem change drastically when using discrete rather than
continuous synapses: optimizing the weights is a convex (and therefore
easy to solve) problem in the latter case, while it is in general
algorithmically hard in the former (NP-complete indeed \cite{amaldi1991complexity}).

The additional computational difficulties associated with discrete
synapses are not however insurmountable: while most traditional approaches
(e.g.~simulated annealing) fail (they scale exponentially with the
problem size, see e.g.~\cite{horner1992dynamics,baldassi_local_2016}),
a number of heuristic algorithms are known to achieve very good performances
even in the extreme case of binary synapses \cite{braunstein-zecchina,baldassi-et-all-pnas,baldassi-2009,baldassi2015max};
some of those are even sufficiently simple and robust to be conceivably
implementable in real neurons \cite{baldassi-et-all-pnas,baldassi-2009}.
The reason why this is at all possible has been an open problem for
some time, because the theoretical analyses on network with binary
synapses consistently described a situation in which even typical
instances should be hard, not only the worst-case ones \cite{huang2014origin}.
In Ref.~\cite{baldassi_subdominant_2015}, we showed that those analyses
were incomplete: the training problem has a huge number of effectively
inaccessible solutions which dominate the statistical measure, but
also dense subdominant regions of solutions which are easily accessed
by the aforementioned heuristic algorithms.

More precisely, the standard statistical analyses of the typical properties
associated with the training problem are concerned with all possible
solutions to the problem, and therefore use a flat measure over all
configurations of synaptic weights that satisfy the training set.
In the binary synapses case, the resulting picture is one in which
with overwhelming probability a solution is isolated from other solutions,
and embedded in a landscape riddled with local minima that trap local
search algorithms like hill climbing or simulated annealing. In contrast,
in our large deviation analysis of Ref.~\cite{baldassi_subdominant_2015},
we have shown that by reweighting the solutions with the number of
other solutions in their neighborhood we could uncover the existence
of extensive regions with very high density of solutions, and that
those regions are the ones which are found by efficient heuristic
algorithms.

In this work, we extend those results from the binary case to the
general discrete case. Our main goal is to show that the above picture
is still relevant in the more biologically relevant scenario in which
synapses can assume more than two states, the sign of the synaptic
weights can not change (also known as Dale's principle), and the inputs
and outputs are generally sparse, as is known to be the case in real
neural networks. To this end, we first extend the standard analysis
(the so-called equilibrium analysis, i.e.~using a flat measure) and
then we repeat the large deviation analysis (i.e.~in which solutions
are reweighted to enhance those of high local density). While this
generalization poses some additional technical difficulties, we find
that, in both cases, the qualitative picture is essentially unaltered:
when the synapses are constrained to assume a limited number of discrete
states, most solutions to the training problem are isolated, but there
exist dense subdominant regions of solutions that are easily accessible.
We also show that the capacity of the network saturates rather fast
with the number of internal synaptic states, and that the aforementioned
accessible regions exist up to a value very close to the network capacity,
suggesting that it may be convenient in a practical implementation
(be it biological or not) to reduce the number of states and instead
exploit the geometrical properties of these dense clusters.

The paper is organized as follows: in Section~\ref{sec:model} we
introduce the discrete perceptron model; in Section~\ref{sec:Equilibrium-analysis}
we compute its capacity as a function of the number of states and
analyze the geometrical properties of typical solutions; in Section~\ref{sec:Large-deviations-analysis}
we describe our large deviations formalism and present the main results
of this paper; in Section~\ref{sec:EdMC} we apply a proof-of-concept
Monte Carlo algorithm driven by the local entropy \cite{baldassi_local_2016}
to our perceptron model; in the Conclusions section we discuss the
scenario emerging from our analysis; in the Appendices we provide
details on the calculations.

\section{Model\label{sec:model}}

We consider a single layer neural network with $N$ inputs and one
output. The network is parametrized by a vector of synaptic weights
$W=\left\{ W_{i}\right\} _{i=1}^{N}$ where each weight can only assume
values from a finite discrete set. For simplicity, throughout this
paper we assume that $W_{i}\in\left\{ 0,1,\dots,L-1,L\right\} $,
but our derivation is general, and holds for arbitrary sets. The network
output is given by
\begin{equation}
\tau\left(W,\xi\right)=\Theta\left(\sum_{i=1}^{N}W_{i}\xi_{i}-\theta N\right)\label{eq:tau}
\end{equation}
where $\xi$ is a vector of inputs of length $N$, $\theta\in\mathbb{R}$
is a neuronal firing threshold, and $\Theta\left(\cdot\right)$ is
the Heaviside step function returning $1$ if its argument is positive
and $0$ otherwise.

We consider training sets composed of $M=\alpha N$ pairs of input/output
associations $\left\{ \xi^{\mu},\sigma^{\mu}\right\} $, $\mu=\left\{ 1,\dots,\alpha N\right\} $,
where $\xi_{i}^{\mu}\in\left\{ 0,1\right\} $ and $\sigma^{\mu}\in\left\{ 0,1\right\} $.
We assume all inputs and outputs to be drawn at random from \emph{biased}
independent identical distributions, with a probability distribution
for each entry given by $P\left(x\right)=\left(1-f\right)\delta\left(x\right)+f\delta\left(x-1\right)$,
where $\delta\left(\cdot\right)$ is the Dirac $\delta$ distribution.
The bias parameter $f$ is often also called \emph{coding rate} in
biological contexts. For simplicity, in the following we will fix
the coding rate $f$ to be the same for the inputs and the outputs,
while in principle we could have chosen two distinct values. 

For any input pattern $\mu$, we can define the error function:
\begin{equation}
E^{\mu}\left(W\right)=\Theta\big(-\left(2\tau\left(W,\xi^{\mu}\right)-1\right)\left(2\sigma^{\mu}-1\right)\big)
\end{equation}
which returns $0$ if the network correctly classifies the pattern,
and $1$ otherwise. Therefore, the training problem of finding an
assignment of weights such that the number of misclassified patterns
is minimized reduces to the optimization of the cost function:
\begin{equation}
E\left(W\right)=\sum_{\mu=1}^{\alpha N}E^{\mu}\left(W\right)\label{eq:energy}
\end{equation}

Finding a configuration for which $E\left(W\right)=0$ is a constraint
satisfaction problem: we denote as 
\begin{equation}
\mathbb{X}_{\xi,\sigma}\left(W\right)=\prod_{\mu=1}^{\alpha N}\left(1-E^{\mu}\left(W\right)\right)\label{eq:indicator}
\end{equation}
the corresponding indicator function. It is generally the case in
these models that, in the limit of large $N$, there is a sharp transition
at a certain value of $\alpha$, called the critical capacity $\alpha_{c}$,
such that the probability (over the distribution of the patterns)
that the problem can be satisfied ($\exists W:\mathbb{X}_{\xi,\sigma}\left(W\right)=1$)
tends to $1$ if $\alpha<\alpha_{c}$ and tends to $0$ if $\alpha>\alpha_{c}$
\footnote{This has not been proved rigorously, but is the result of non-rigorous
replica theory analysis and numerical simulations supporting the statement.
For rigorous works providing bounds to the transitions, see~\cite{kim_covering_1998,talagrand_intersecting_1999}.}. Some well-known values of $\alpha_{c}$ in similar models are $\alpha_{c}=2$
in the case of unbounded continuous weights and unbiased inputs in
$\left\{ -1,1\right\} $ \cite{gardner-derrida}; $\alpha_{c}=1$
in the same situation but with positive continuous weights and inputs
in $\left\{ 0,1\right\} $ (this also corresponds to the limiting
case $L\to\infty$ of the model of eq.~\eqref{eq:tau}) \cite{Brunel2004};
$\alpha_{c}=0.833$ in the case of both inputs and weights taking
values in $\left\{ -1,+1\right\} $ with unbiased inputs \cite{krauth-mezard};
and $\alpha_{c}=0.59$ for the model of eq.~\eqref{eq:tau} for the
binary case $L=1$ and unbiased inputs $f=0.5$ \cite{gutfreund1990capacity}.
In all these cases, the neuronal threshold $\theta$ needs to be chosen
optimally in order to maximize the capacity: for example, in the latter
case of $L=1$ and $f=0.5$ the optimal value is $\theta\simeq0.16$.
In the next section we will show the values of $\alpha_{c}$ for general
$L$ and $f$.

The choice of using uncorrelated inputs and outputs is arguably not
very realistic, both from the point of view of biological modeling
and for machine learning applications. This simple scenario is also
known as a \emph{classification} task; it is possible to consider
instead the case where the outputs $\sigma^{\mu}$ are modeled as
being produced from some underlying rule which the device has to discover
from the training set, the so called \emph{generalization} task. The
latter is certainly a more relevant scenario for many applications.
Nevertheless, in the binary case of our previous work~\cite{baldassi_subdominant_2015}
we showed that these assumptions -- which were taken in order to simplify
the theoretical analysis -- seem to leave the resulting qualitative
picture unaltered, and therefore we argue that it is rather likely
that the situation would be similar for the multi-valued model studied
in this paper. We will come back to this issue in the discussion of
Section \ref{sec:Large-deviations-analysis}.

\section{Equilibrium analysis\label{sec:Equilibrium-analysis}}

\subsection{Critical capacity as a function of the number of synaptic states\label{sec:Critical-capacity-as}}

As a first step towards extending our large deviation analysis to
the model described by eq.~\eqref{eq:tau}, we performed a standard
equilibrium analysis and verified that the scenario is the same that
holds in other similar models. We could also use this analysis to
compute the theoretical critical capacity of the system as a function
of the number of states per synapse $L+1$ and of the coding rate
$f$.

This kind of analysis, often called \emph{à la Gardner} \cite{gardner-derrida},
consists in studying the typical thermodynamical properties of a system
described by the Boltzmann measure
\begin{equation}
P\left(W;\beta\right)=\frac{1}{Z}e^{-\beta E\left(W\right)}
\end{equation}
where $E\left(W\right)$ is defined in eq.~\eqref{eq:energy} and
$Z$ (also known as the partition function) is a normalization constant,
in the zero-temperature limit $\beta\to\infty$. This is therefore
a flat measure on the ground states of the system. When perfect learning
is possible, i.e.~$\min_{W}E\left(W\right)=0$, we have (see eq.~\eqref{eq:indicator}):
\begin{equation}
P_{F}\left(W\right)=\frac{\mathbb{X}_{\xi,\sigma}\left(W\right)}{\sum_{W^{\prime}}\mathbb{X}_{\xi,\sigma}\left(W^{\prime}\right)}\label{eq:flat_measure}
\end{equation}
where the subscript ``$F$'' stands for ``flat''. In order to
describe the typical behavior of a system we need to compute the average
over the patterns of the entropy density:
\begin{eqnarray}
\Phi & = & \frac{1}{N}\left\langle \log\sum_{W}\mathbb{X}_{\xi,\sigma}\left(W\right)\right\rangle 
\end{eqnarray}
where $\left\langle \cdot\right\rangle $ denotes the average over
the distribution of the patterns. This computation is accomplished
by the so-called replica trick and, although not rigorous, is believed
to provide the correct values for some relevant quantities such as
the optimal value of the neuronal threshold $\theta$ and the critical
capacity, which in this case is derived as the value of $\alpha$
for which $\Phi=0$.

\begin{figure}
\includegraphics[width=1\textwidth]{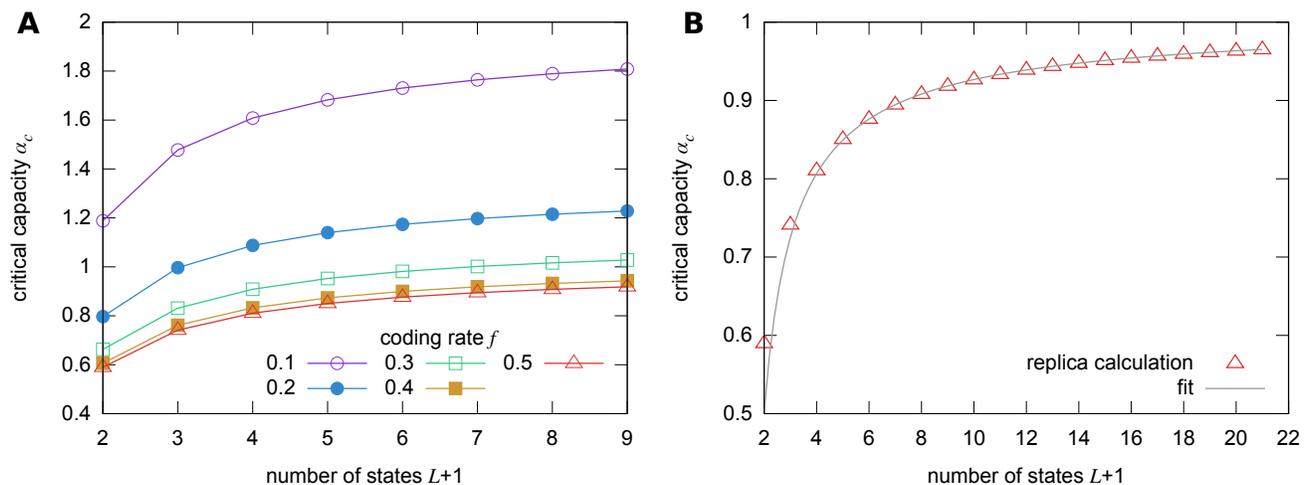}

\protect\caption{\label{fig:capacity_vs_numstates}\textbf{A.} Critical capacity $\alpha_{c}$
as a function of the number of states per synapse $L+1$, for different
values of the coding rate $f$. \textbf{B.} Same as in panel A, but
only for the dense (unbiased) case $f=0.5$, with a wider range of
$L$, and showing a fit of the form $\alpha^{\infty}-\frac{a}{L^{b}}$
over the last part of the curve ($L\ge5$). The fitted parameters
are $\alpha^{\infty}\simeq1.0$, $a\simeq0.5$, $b=0.85$.}
\end{figure}

The details of the computation follow standard steps (they can also
be obtained from the computation presented in Appendix~\ref{AppendixB:Modified-measure}
setting $y=0$). Fig~\ref{fig:capacity_vs_numstates}A shows the
resulting value of $\alpha_{c}$ as a function of the number of states
$L+1$, for different values of the coding rate $f$. As expected,
$\alpha_{c}$ increases with the number of values a synaptic variable
can assume and with the sparsity of the coding. Fig.~\ref{fig:capacity_vs_numstates}B
shows the same curve for the dense (unbiased) $f=0.5$ case with a
wider range of $L$: it is expected that in this case $\alpha_{c}\to1$
as $L\to\infty$, consistently with the case of continuous positive
synapses \cite{Brunel2004}, and therefore we also show the results
of a tentative fit of the form $\alpha_{c}\sim\alpha^{\infty}-\frac{a}{L^{b}}$
which estimates the rate of convergence to the continuous case; the
fit yields $\alpha^{\infty}\simeq1.0$, as expected, and an exponent
$b\simeq0.85$. From the results in Fig.~\ref{fig:capacity_vs_numstates}A,
it can be seen that the qualitative behavior is not different for
the sparser cases. Qualitatively similar results were also obtained
in a slightly different setting in~\cite{gutfreund1990capacity}.

One interesting general observation about these results is that the
gain in capacity with each additional synaptic state decreases fairly
rapidly after the first few values. This observation by itself is
not conclusive, since even when solutions exist they may be hard to
find algorithmically (see the next section). As we shall see in Section~\ref{sub:alpha_U_vs_states},
however, \emph{accessible} solutions exist for all the cases we tested
at least up to $0.9\alpha_{c}$. From the point of view of the implementation
cost (whether biological or in silico), it seems therefore that increasing
the synaptic precision would not be a sensible strategy, as it leads
to a very small advantage in terms of computational or representational
power. This is consistent with the general idea that biological synapses
would only need to implement synapses with a few bits of precision.

\subsection{Typical solutions are isolated}

To explore the solution space structure of a perceptron learning problem
the general idea is to select a reference solution sampled from the
flat measure of eq.~\eqref{eq:flat_measure}, and count how many
other solutions can be found at a given distance from the selected
one. This technique is known as the Franz-Parisi potential \cite{franz1995recipes}.

First, we define the \emph{local entropy} density for a given reference
configuration $\tilde{W}$ at a given distance $D$ as

\begin{equation}
\mathcal{S}_{\xi,\sigma}\left(\tilde{W},D\right)=\frac{1}{N}\log\sum_{\left\{ W\right\} }\mathbb{X}_{\xi,\sigma}\left(W\right)\delta\left(d\left(W,\tilde{W}\right)-D\right)\label{eq:local_entropy}
\end{equation}
i.e.~as the logarithm of the number of solutions at distance $D$
from $\tilde{W}$, having defined the $d\left(W,\tilde{W}\right)$
as a normalized distance function:

\begin{equation}
d\left(W,\tilde{W}\right)=\frac{1}{4N}\sum_{i}\left(W_{i}-\tilde{W}_{i}\right)^{2}\label{eq:distance}
\end{equation}

We introduced the factor $\nicefrac{1}{4}$ for consistency with the
computation in~\cite{baldassi_subdominant_2015}: in the case where
$W_{i}\in\left\{ -1,+1\right\} $, this reduces to the Hamming distance.

Sampling from a Boltzmann distribution means looking at the \emph{typical}
structure of the solution space, i.e.~computing the typical \emph{local
entropy} density:
\begin{equation}
\mathcal{S}_{FP}\left(D\right)=\left\langle \sum_{\left\{ \tilde{W}\right\} }P_{F}\left(\tilde{W}\right)\mathcal{S}_{\xi,\sigma}\left(\tilde{W},D\right)\right\rangle \label{eq:typical_local_entropy}
\end{equation}
where the subscript ``$FP$'' stands for Franz-Parisi, $\tilde{W}$
represents the reference equilibrium solution, and $\left\langle \cdot\right\rangle $
as usual is the average over the disorder (the patterns). 

Again, the computation can be performed with the replica method, and
is detailed in Appendix~\ref{sec:AppendixA-FP}. The results of this
typical case analysis are shown as the black lines in Fig.~\ref{fig:local_entropy_vs_distance}
and are qualitatively the same as those already obtained in models
with binary synapses, regardless of the number of synaptic states
or the coding rate: namely, for all values of the parameters -- and
in particular, for all $\alpha>0$ -- there exist a value $D_{\textrm{min}}$
such that for $D\in\left(0,D_{\textrm{min}}\right)$ we obtain $\mathcal{S}_{FP}\left(D\right)<0$.
This (unphysical) result is assumed to signal the onset of replica
symmetry breaking effects and that the actual value of the local entropy
is $0$, which means that typical solutions are isolated in the space
of configurations, when considering neighborhoods whose diameter is
of order $N$.

Isolated solutions are very hard to find algorithmically. As we have
shown in~\cite{baldassi_subdominant_2015}, the efficient algorithms,
i.e.~which experimentally exhibit sub-exponential scaling in computational
complexity with $N$, find non-isolated solutions, which are therefore
not typical. In the next section, we will show that these subdominant
solutions exist also in the multi-valued model we are considering
in the present work.

\section{Large deviations analysis\label{sec:Large-deviations-analysis}}

Following~\cite{baldassi_subdominant_2015}, we introduce a large
deviation measure in order to describe regions of the configuration
space where the solutions to the training set are maximally locally
dense. To this end, we modify the flat distribution of eq.~\eqref{eq:flat_measure}
by increasing the relative weight of the solutions with a higher local
entropy (eq.~\ref{eq:local_entropy}), as follows: 
\begin{equation}
P_{RC}\left(\tilde{W};y,D\right)=\frac{\mathbb{X}_{\xi,\sigma}\left(\tilde{W}\right)e^{yN\mathcal{S}_{\xi,\sigma}\left(\tilde{W},D\right)}}{\sum_{\tilde{W}^{\prime}}\mathbb{X}_{\xi,\sigma}\left(\tilde{W}^{\prime}\right)e^{yN\mathcal{S}_{\xi,\sigma}\left(\tilde{W}^{\prime},D\right)}}
\end{equation}
where the subscript ``$RC$'' stands for ``reweighted, constrained''.
The parameter $y$ has the role of an inverse temperature: by taking
the limit $y\to\infty$ this distribution describes the solutions
of maximal local density. 

Alternatively, we can just use $\tilde{W}$ as a reference configuration
without enforcing the constraint $\mathbb{X}_{\xi,\sigma}\left(\tilde{W}\right)$,
and obtain:
\begin{equation}
P_{RU}\left(\tilde{W};y,D\right)=\frac{e^{y\mathcal{S}_{\xi,\sigma}\left(\tilde{W},D\right)}}{\sum_{\tilde{W}^{\prime}}e^{y\mathcal{S}_{\xi,\sigma}\left(\tilde{W}^{\prime},D\right)}}
\end{equation}
where the subscript ``$RU$'' stands for ``reweighted, unconstrained''.

We can study the typical behavior of these modified measures as usual
within the replica theory, by computing their corresponding average
free entropy density:
\begin{eqnarray}
\Phi_{RC}\left(D,y\right) & = & \frac{1}{N}\left\langle \log\sum_{\tilde{W}}\mathbb{X}_{\xi,\sigma}\left(\tilde{W}\right)e^{y\mathcal{S}_{\xi,\sigma}\left(\tilde{W},D\right)}\right\rangle \label{eq:free_entropy_RC}\\
\Phi_{RU}\left(D,y\right) & = & \frac{1}{N}\left\langle \log\sum_{\tilde{W}}e^{y\mathcal{S}_{\xi,\sigma}\left(\tilde{W},D\right)}\right\rangle \label{eq:free_entropy_RU}
\end{eqnarray}

With these, we can compute the typical values of the local entropy
density
\begin{equation}
\mathcal{S}_{RC}\left(D,y\right)=\frac{\partial}{\partial y}\Phi_{RC}\left(D,y\right)
\end{equation}
and of the external entropy density
\begin{equation}
\Sigma_{RC}\left(D,y\right)=\Phi_{RC}\left(D,y\right)-y\mathcal{S}_{RC}\left(D,y\right)\label{eq:ext_entropy}
\end{equation}
(analogous relations hold for the $RU$ case).

The latter quantity measures the logarithm of the number of reference
configurations $\tilde{W}$ that correspond to the given parameters
$y$ and $D$, divided by $N$. Since the systems are discrete, both
these quantities need to be non-negative in order for them to represent
typical instances, otherwise they can only be interpreted in terms
of rare events \cite{rivoire_properties_2004}.

There are several reasons for studying both the $RC$ and the $RU$
cases. The $RC$ case is more directly comparable with the typical
$FP$ case: as such, it is the most straightforward way to demonstrate
that the large deviations analysis paints a radically different picture
about the nature of the solutions than the equilibrium case. Furthermore,
when studying the problem at finite $y$, only the constrained case
gives reasonable results when assuming replica symmetry (see below).
The $RU$ case, on the other hand, can be exploited in designing search
algorithms (Sec.~\ref{sec:EdMC}); moreover, as explained below,
the $RC$ case reduces to the $RU$ case in the limit $y\to\infty$.
Finally, since both cases are problematic, due to the numerical difficulties
in solving the saddle point equations and to the possible presence
of further levels of replica symmetry breaking, the accuracy of the
results may be questioned. Their comparison however shows that the
results of the different analyses are in quite good agreement: this
observation, complemented by numerical experiments, provides an indication
that the results are reasonably accurate.

\subsection{Reweighted Constrained distribution, RS analysis}

In the case of the computation of $\Phi_{RC}\left(D,y\right)$, eq.~\eqref{eq:free_entropy_RC},
we performed the analysis using a replica-symmetric (RS) Ansatz. The
details are provided in Appendix~\ref{AppendixB:Modified-measure}.
Since we are interested in the configurations of highest density,
we want to take the parameter $y$ to be as high as possible, in principle
we wish to study the case $y\to\infty$. However, in this case the
external entropy $\Sigma_{RC}\left(D,y\right)$ is negative for all
values of the parameters. This signals a problem with the RS Ansatz,
and implies that we should instead consider replica-symmetry-broken
solutions. In geometrical terms, the interpretation is as follows:
the RS solution at $y\to\infty$ implies that the typical overlap
between two different reference solutions $\tilde{W}^{a}$ and $\tilde{W}^{b}$,
as computed by $\tilde{q}=\frac{1}{N}\sum_{i}\tilde{W}_{i}^{a}\tilde{W}_{i}^{b}$,
tends to $\tilde{Q}=\frac{1}{N}\sum_{i}\tilde{W}_{i}^{a}\tilde{W}_{i}^{a}$
(see Sec.~\ref{sec:RS-solution-large-y}), and therefore that there
should be a single solution of maximal local entropy density. The
fact that the RS assumption is wrong implies that the structure of
the configurations of maximal density is more complex, and that, at
least beyond a certain $y$, the geometry of the reference configurations
$\tilde{W}$ breaks into several clusters (see comments at the end
of the following section).

Because of technical issues in solving the equations at the 1RSB level
(namely, the fact that the resulting system of equations is too large
and that some of the equations involve multiple nested integrals that
are too expensive to compute in reasonable times for arbitrary $y$),
we used instead the maximum value of $y$ for which the RS results
are physically meaningful, as we did already in~\cite{baldassi_subdominant_2015}.
Therefore, for any given $\alpha$, we computed $y^{\star}\left(D\right)$
such that $\Sigma_{RC}\left(D,y^{\star}\left(D\right)\right)=0$.

The 1RSB equations simplify in the limit $y\to\infty$, but that still
does not solve the problem of the negative external entropy, suggesting
that the correct solution requires further levels of replica symmetry
breaking. We come back to this point in the next section~\eqref{sub:RU},
where we also comment on the results of the analysis, shown in Fig.~\ref{fig:local_entropy_vs_distance}.

\subsection{Reweighted Unconstrained distribution, 1RSB analysis\label{sub:RU}}

The unconstrained case, $\Phi_{RU}\left(D,y\right)$, eq.~\eqref{eq:free_entropy_RU},
is considerably simpler. However, replica symmetry breaking effects
are also stronger, leading to clearly unphysical results at the RS
level even when using $y^{\star}\left(D\right)$ such that $\Sigma_{RU}\left(D,y^{\star}\left(D\right)\right)=0$
(for example, this solution would predict a positive local entropy
for some values of the parameters even beyond $\alpha_{c}$, which
does not make sense).

Therefore, this case needs to be studied at least at the level of
1RSB. The details are provided in Appendix~\ref{sec:AppendixD-1RSB}.
Again, the resulting equations are computationally still very heavy,
and we could not explore the whole range of parameters at finite $y$.
In the limiting case $y\to\infty$ the equations simplify and the
computational complexity is comparable to the constrained case at
finite $y$ in the RS Ansatz. Interestingly, in this limit the thermodynamic
quantities are identical in the constrained and unconstrained case.

This solution does not solve the problem of negative external entropy,
implying that further levels of replica symmetry breaking are required,
but the situation improves considerably: the unphysical branches beyond
$\alpha_{c}$ disappear, and the modulus of the external entropy is
very small and tends to $0$ as $D\to0$. Furthermore, the results
of the 1RSB analysis at $y\to\infty$ and of the RS analysis of the
constrained case at $y=y^{\star}\left(D\right)$ are qualitatively
essentially the same and quantitatively very close, which suggests
that these results provide a good approximation to the description
of the regions of highest local entropy density in the configuration
space. Furthermore, all the results are completely analogous to the
ones obtained in the binary balanced unbiased case (the constrained
RS analysis was shown in~\cite{baldassi_subdominant_2015} and the
1RSB analysis in~\cite{baldassi_local_2016}), where it was also
shown that numerical experiments, where available, match very closely
the theoretical predictions.

\begin{figure}
\includegraphics[width=1\textwidth]{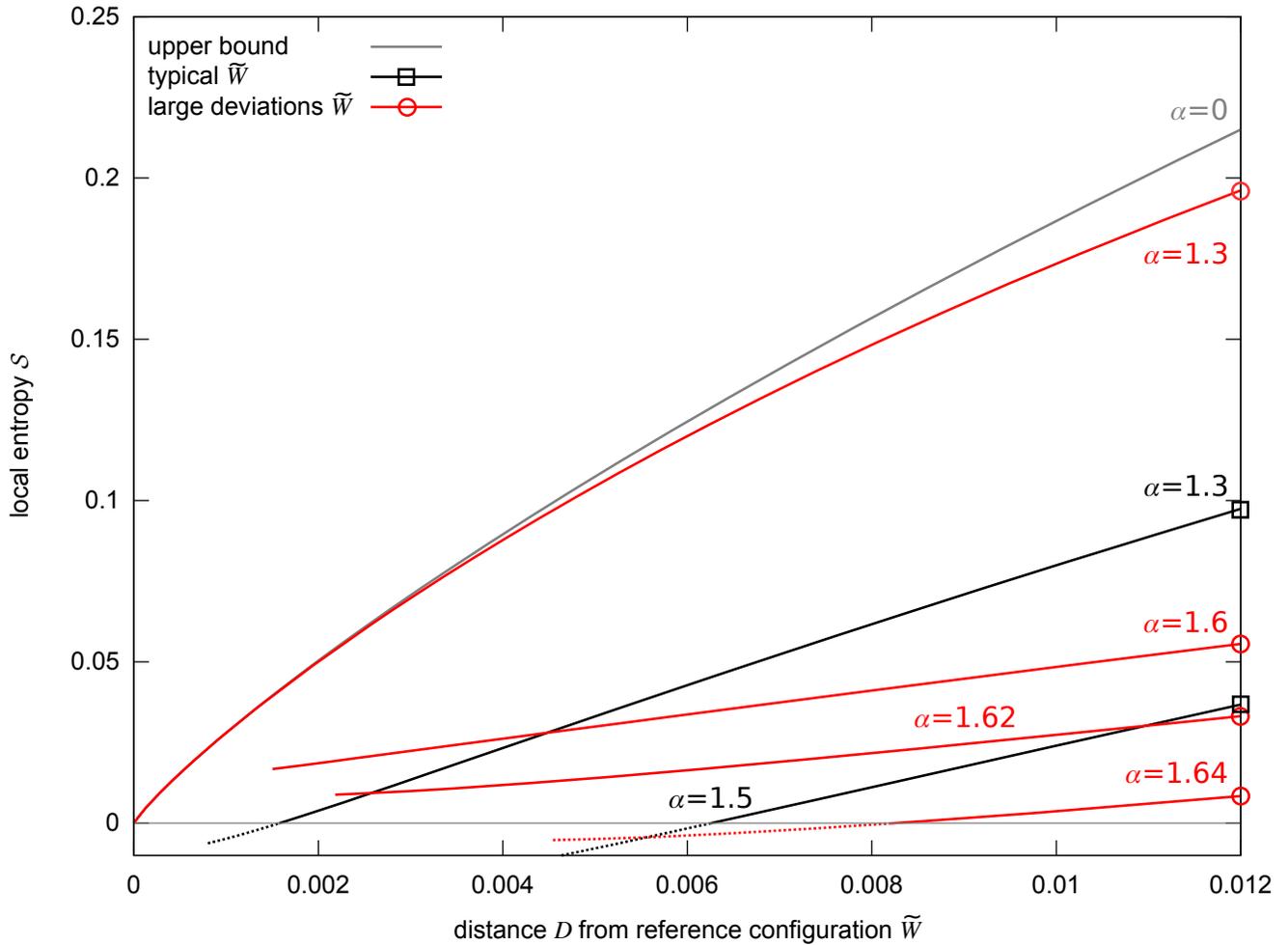}

\protect\caption{\label{fig:local_entropy_vs_distance}Local entropy density as a function
of the distance $D$ from the reference configuration $\tilde{W}$,
comparing the typical case from the Franz-Parisi analysis (black lines,
marked with squares) with the large deviations case (red lines, marked
with circles), at various values of the number of patterns per variable
$\alpha$. The upper bound (gray dashed curve) corresponds to the
$\alpha=0$ case where every configuration is a solution. The unphysical
portions of the curves where the local entropy becomes negative is
dotted. For the typical case, all curves eventually go below zero
at some $D_{min}>0$, for all values of $\alpha$, i.e.~typical solutions
are isolated. For the large deviations case, the curves for the $\Phi_{RC}\left(D,y^{\star}\left(D\right)\right)$
case (RS analysis) and the $\Phi_{RU}\left(D,\infty\right)$ case
(1RSB analysis) yield results which are too close to be distinguished
in the plot at this resolution. The ``large deviations $\tilde{W}$''
curve at $\alpha=1.6$ is interrupted due to numerical problems in
solving the equations, but it could continue up to $D=0$, approaching
the upper bound for small $\alpha$. Our results indicate that that
is the case for $\alpha=1.55$, although it is not shown here since
we could not produce a complete curve, again due to numerical difficulties.
The curves for $\alpha=1.62$ and $\alpha=1.64$ are interrupted because
the equations stop having solutions at some value of $D>0$ ($\alpha_{U}$
transition, see text). The large deviations curve at $\alpha=1.3$
is also essentially indistinguishable from the RS computation performed
at $y=\infty$.}
\end{figure}

In Fig.~\ref{fig:local_entropy_vs_distance} we show the predictions
for the local entropy as a function of the distance in one representative
case, for $L=4$ and $f=0.1$, for different values of $\alpha$,
for the three cases: $\mathcal{S}_{FP}\left(D\right)$ (eq.~\eqref{eq:typical_local_entropy}),
$\mathcal{S}_{RC}\left(D,y^{\star}\left(D\right)\right)$ (derived
from eq.~\eqref{eq:free_entropy_RC}) and $\mathcal{S}_{RU}\left(D,\infty\right)$
(derived from eq.~\eqref{eq:free_entropy_RU}). The latter two cases
give results which, where it was possible to directly compare them,
are quantitatively so close that the difference cannot be appreciated
at the resolution level of the plotted figure, and thus we treat the
two cases as equivalent for the purposes of the description of the
results. In both those cases, the solution of the equations become
numerically extremely challenging around the transition point $\alpha_{U}$
(see below for the definition) and thus we could not complete all
the curves in that region. The most notable features that emerge from
this figure are:
\begin{itemize}
\item Typical solutions are isolated: $\mathcal{S}_{FP}\left(D\right)$
becomes negative in a neighborhood of $D=0$.
\item Up to a certain $\alpha_{U}<\alpha_{c}$ (where $\alpha_{U}$ is between
$1.55$ and $1.62$ for the specific case of Fig.~\ref{fig:local_entropy_vs_distance}),
there exist dense regions of solutions: in this phase, there exist
non-typical solutions that are surrounded by an exponential (in $N$)
number of other solutions, and at small distances the local entropy
curves tend to collapse onto the $\alpha=0$ curve, which corresponds
to the upper bound where each configuration is a solution.
\item Between $\alpha_{U}$ and $\alpha_{c}$, there are regions of $D$
where either there is no solution to the equations or the solution
leads to a negative local entropy; in both cases, we interpret these
facts as indicating a change in the structure of the dense clusters
of solutions, which either disappear or break into small disconnected
and isolated components.
\end{itemize}
The significance of the phase transition at $\alpha_{U}$ is related
to the accessibility of the dense regions of solutions and the existence
of efficient algorithms that are able to solve the training task.
In the case of the binary, balanced and unbiased case studied in~\cite{baldassi_subdominant_2015},
our best estimate was $\alpha_{U}\simeq0.76$, while the best available
heuristic algorithms (Belief Propagation with reinforcement \cite{braunstein-zecchina},
Max-Sum with reinforcement \cite{baldassi2015max}) were measured
experimentally to have a capacity of $0.75$ and the theoretical critical
capacity is believed to be $\alpha_{c}=0.83$ \cite{krauth-mezard}.
Another (simpler but faster) heuristic algorithm, called SBPI, was
measured to achieve a slightly lower capacity, reaching almost $\alpha=0.7$
\cite{baldassi-et-all-pnas}. A very similar situation happens with
the same model in the generalization scenario, where $\alpha_{U}\simeq1.1$
\cite{baldassi_subdominant_2015} is very close to the maximum value
reached by the best heuristic solvers \cite{baldassi2015max}, leaving
a region where the heuristics fail before the theoretical transition
to perfect learning at $1.25$ \cite{sompolinsky1990learning}. For
the dense binary case with $W_{i}\in\left\{ 0,1\right\} $, i.e.~the
model considered in this paper with $L=1$ and $f=0.5$, SBPI was
measured to achieve a capacity slightly above $\alpha\simeq0.5$ \cite{baldassi-et-all-pnas},
to be compared to the theoretical maximum $\alpha_{c}=0.59$ \cite{gutfreund1990capacity}.
For this case, the large deviation analysis gives $\alpha_{U}\simeq0.54$.
It was also shown by direct numerical experiments in~\cite{baldassi_subdominant_2015}
that all solutions found by the heuristic algorithms at sufficiently
large $N$ are part of a dense region that is well described by the
large deviation analysis.

All these results thus strongly suggest that $\alpha_{U}$ signals
a transition between an ``easy'' phase and a ``hard'' phase. This
situation bears some clear similarities with other constraint satisfaction
problems like random $K$-satisfiability ($K$-SAT), where in particular
there can be a ``frozen'' phase where solutions are isolated and
no efficient algorithms are known to work \cite{krzakala-csp}. Contrary
to the $K$-SAT case, however, in the case of neural networks this
transition does not appear in the equilibrium analysis -- which would
predict that the problem is intractable at all values of $\alpha$
-- but only in a large deviations study. This latter observation is
presumably linked with the complex geometrical structure of the dense
regions, which are not ``states'' in the usual sense given to the
word in the context of statistical physics of complex systems, i.e.~they
are not clearly separated clusters of configurations, according to
the argument that otherwise it should not have been necessary to perform
the large deviation analysis in the first place in order to observe
them. Our analysis (theoretical and numerical) is not sufficient to
completely characterize this geometrical structure, apart from telling
us that it must be extensive, that the density seems to vary in a
continuous fashion (i.e.~the local entropy landscape is rather smooth,
such that it is algorithmically easy to find a path towards a solution,
see Sec.~\ref{sec:EdMC}), and that there are several (but less than
exponentially many) regions of highest density (due to replica symmetry
breaking effects, i.e.~not related to any obvious symmetry of the
problem). These highest density regions are, to the leading exponential
order, all equivalent, thus a local search algorithm designed to exploit
their existence needs to be able to spontaneously break the symmetry
among them. It would indeed be very interesting to be able to further
refine this description.

\subsection{Transition point $\alpha_{U}$ as a function of the number of states\label{sub:alpha_U_vs_states}}

Determining the value of $\alpha_{U}$, where the dense regions seem
to disappear (or are at least no longer easily accessible), is extremely
challenging computationally, not only because of the time-consuming
task of solving the system of equations that result from the replica
analysis (and which require repeated nested numerical integrations),
but especially due to purely numerical issues related to the finite
machine precision available and the trade-offs involved between computational
time and increased precision. These issues are exacerbated near the
transition point.

However, despite the fact that the RS analysis in the limit $y\to\infty$
(performed in Appendix~\ref{sec:RS-solution-large-y}) gives some
unphysical results that need to be corrected at higher level of symmetry
breaking, it still provides an estimate of $\alpha_{U}$, which can
be computed reasonably efficiently, and which is not dramatically
affected by the RSB corrections. For example, in the binary, balanced
unbiased case of~\cite{baldassi_subdominant_2015}, the RS analysis
at $y\to\infty$ gives $\alpha_{U}\simeq0.755$, while the 1RSB solution
gives $\alpha_{U}\simeq0.76$. In the case of the multi valued model
of this paper with the parameters $L=4$ and $f=0.1$ used for Fig.~\ref{fig:local_entropy_vs_distance},
the RS analysis at $y\to\infty$ gives $\alpha_{U}\simeq1.6$ while
the 1RSB analysis gives $\alpha_{U}$ between $1.55$ and $1.62$.

\begin{figure}
\includegraphics[width=1\textwidth]{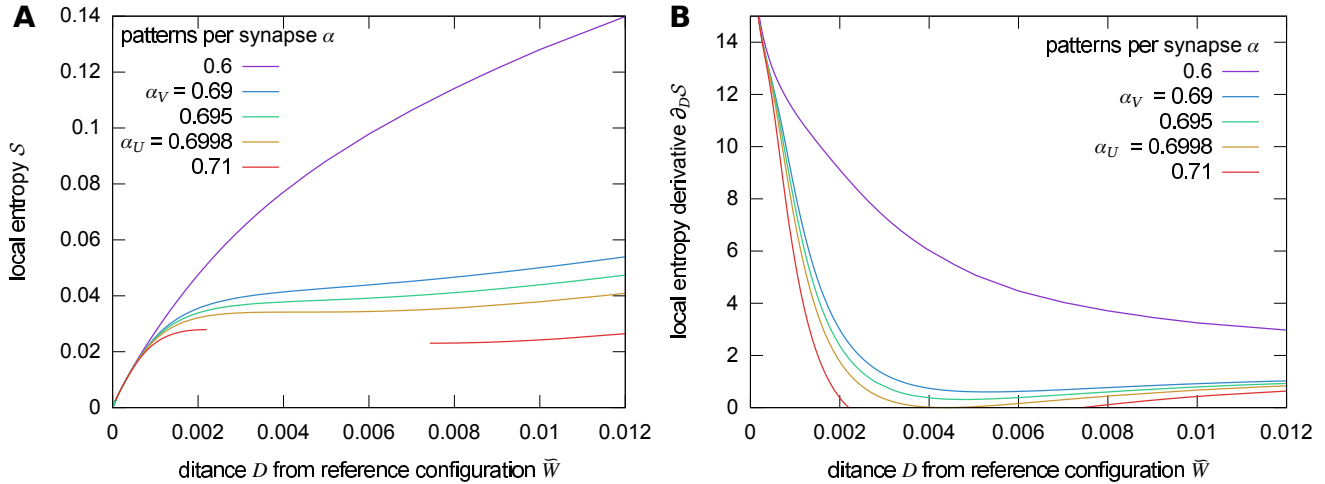}

\protect\caption{\label{fig:nentropy}\textbf{A.} Local entropy density as a function
of the distance from the reference solution $\tilde{W}$, for $L=3$
and $f=0.5$, in the approximation of RS at $y\to\infty$, for various
values of $\alpha$, showing five representative curves (from top
to bottom): $\alpha<\alpha_{V}$, $\alpha=\alpha_{V}=0.69$, $\alpha_{V}<\alpha<\alpha_{U}$,
$\alpha=\alpha_{U}=0.6998$, $\alpha>\alpha_{U}$. \textbf{B.} Derivative
of the local entropy with respect to the distance $D$, for the same
case as for panel A. Curves are still arranged from top to bottom.
This shows the change in concavity occurring at $\alpha_{V}$ and
the gap appearing at $\alpha_{U}$.}
\end{figure}

Therefore, we have used the RS analysis at $y\to\infty$ (note that
in this limit there is no difference between the constrained free
entropy $\Phi_{RC}$ and the unconstrained free entropy $\Phi_{RU}$,
see the discussion in Appendix~\ref{sec:RS-solution-large-y}) to
explore the behavior of $\alpha_{U}$ when varying the number of states
and the coding level of the patterns. This is most easily achieved
by studying the derivative of the local entropy as a function of the
distance $\partial_{D}\mathcal{S}\left(D,\infty\right)$: Fig\@.~\ref{fig:nentropy}A
shows an example of the behavior of the local entropy as a function
of the distance for various values of $\alpha$ in the dense ternary
case $L=2$, $f=0.5$.

As one can notice, there are three types of behavior: (i) below a
certain $\alpha_{V}$ the local entropy curves are concave; (ii) between
$\alpha_{V}$ and $\alpha_{U}$ there appear intermediate regions
where the curve becomes convex; (iii) between $\alpha_{U}$ and $\alpha_{C}$
a gap appears, where there are no solutions. The appearance of the
gap (and thus $\alpha_{U}$) is signaled by the fact that it is the
lowest value of $\alpha$ for which there exists a $D$ such that
$\partial_{D}\mathcal{S}\left(D,\infty\right)=0$ (Fig.~\ref{fig:nentropy}B).
These qualitative observations remain unchanged in the 1RSB case and
across all neural network models we have studied.

The appearance of the gap at $\alpha_{U}$ seems to be related to
a breaking apart of the structure of the dense regions, which we also
observed numerically at relatively small $N$. It is not completely
clear whether the branch after the break is physical or an artifact
of the replica analysis, since we are unable --- for the time being
--- to find such regions numerically at large $N$, and thus to confirm
their existence. If it is physical, then it depicts a situation in
which dense regions still exist, but are broken apart into several
separated clusters and are no longer as easily accessible as for $\alpha<\alpha_{U}$.

\begin{figure}
\includegraphics[width=1\textwidth]{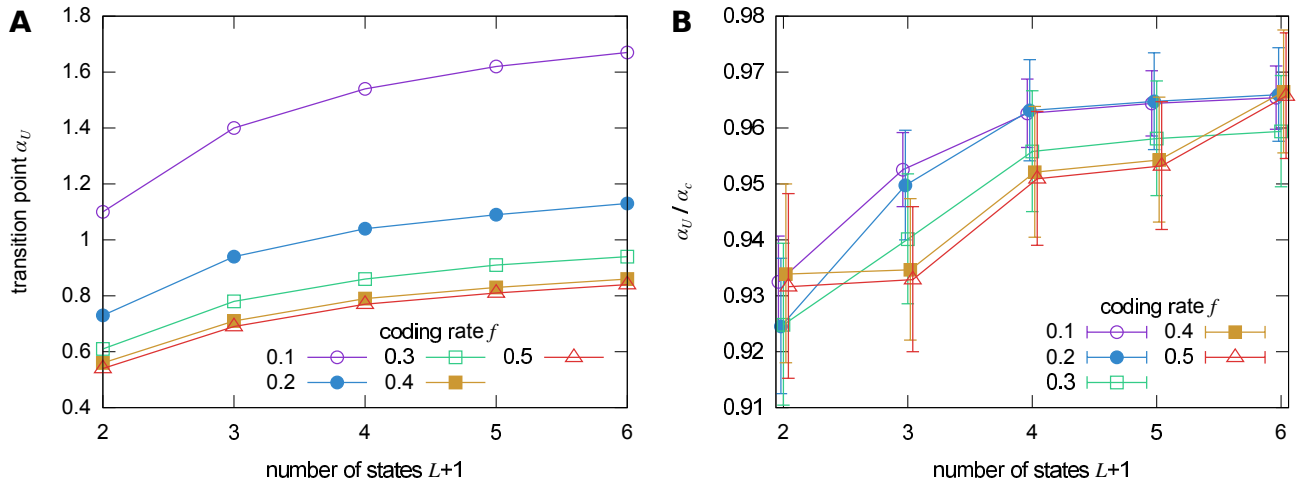}

\protect\caption{\label{fig:alpha_U_vs_numstates}\textbf{A.} Transition point $\alpha_{U}$
as a function of the number of states per synapse $L+1$, for different
values of the coding rate $f$, as computed in the approximation of
RS at $y\to\infty$. \textbf{B.} Same as panel A, but $\alpha_{U}$
is divided by the critical capacity $\alpha_{c}$. Error bars reflect
the finite precision in the determination of the values. Points for
different values of $f$ are slightly shifted relative to each other
for improved legibility. Despite the limited number of values, a general
tendency of this value to increase with $L$ is observed (the ratio
is expected to tend to $1$ for $L\to\infty$), while the dependency
on $f$ is less clear.}
\end{figure}

The behavior of $\alpha_{U}$ as a function of the number of states,
for various values of the coding level $f$, is shown in Fig.~\ref{fig:alpha_U_vs_numstates}A.
The behavior is very close to that of the critical capacity, cf.~Fig.~\ref{fig:capacity_vs_numstates}.
We expect that these quantities converge to the same value in the
limit $L\to\infty$ where the device should behave as in the case
of continuous synapses, which seem to be the case, see Fig.~\ref{fig:alpha_U_vs_numstates}B.

\section{Proof of concept: generalizing Entropy-driven Monte Carlo\label{sec:EdMC}}

The existence of subdominant dense clusters of solutions not only
serves to provide a plausible explanation for the observed behavior
of existing heuristic algorithms: it can also be exploited to design
new algorithms. As a proof of concept, in~\cite{baldassi_local_2016},
we have presented an algorithm called ``Entropy-driven Monte Carlo''
(EdMC), that exploits the fact that the landscape of the local entropy
can be radically different from that of the energy.

The basic idea is to run a Simulated Annealing (SA) algorithm using
the local entropy as an objective function rather than the energy,
as follows: at any given configuration $\tilde{W}$, we consider a
nearby configuration $\tilde{W}^{\prime}$ (obtained by picking uniformly
at random a synaptic index $i$ and then randomly increasing or decreasing
$\tilde{W_{i}}$ by one) and estimate the shift in local entropy $\mathcal{S}_{\xi,\sigma}\left(\tilde{W}^{\prime},D\right)-\mathcal{S}_{\xi,\sigma}\left(\tilde{W},D\right)$
(see eq.~\eqref{eq:local_entropy}), and accept or reject the move
$\tilde{W}\to\tilde{W}^{\prime}$ according to the Metropolis rule
at an inverse temperature $y$. After a number of accepted moves,
we increase $y$ and reduce $D$ by a fixed amount, until we eventually
find a solution. We call the process of gradually reducing $D$ ``scoping'',
in analogy with the ``annealing'' process of increasing $y$.

The estimation of the local entropy is performed using Belief Propagation
(BP) \cite{mackay2003information,yedidia2005constructing}; for simplicity,
instead of imposing a hard constraint on the distance $D$, we alternatively
fix the value of its Legendre conjugate parameter by introducing a
collection of fixed external fields (of a defined intensity $\gamma$)
in the direction of $\tilde{W}$, as described in detail in~\cite{baldassi_local_2016}.
The scoping process is thus obtained by gradually increasing $\gamma$.

The tests performed with this algorithm show that, while standard
Simulated Annealing using the energy $E\left(\tilde{W}\right)$ (the
number of misclassified patterns, see eq.~\eqref{eq:energy}) as
the objective function gets immediately trapped by the exponentially
large number of local minima, EdMC does not, and can reach a solution
even in the greedy case in which it is run directly at zero temperature
($y\to\infty$).

While this algorithm is certainly slower than other efficient heuristic
solvers, it is still interesting for these reasons: (i) it is generic,
since it can in principle be generalized to any model where reasonable
estimates of the local entropy can be achieved; (ii) it is more ``under
control'' than the heuristic alternatives, since its behavior actually
closely matches the theoretical prediction of the large deviation
analysis; (iii) it proves that the local entropy landscape is very
different from the energy landscape (and EdMC could obviously be used
directly to explore such a landscape, if run as a simple Monte Carlo
algorithm without scoping or annealing).

In any case, it is also easy to heuristically improve this algorithm
dramatically, by using the BP fixed point messages to propose the
moves, rather then performing them at random (but still using $\mathcal{S}_{\xi,\sigma}\left(\tilde{W},D\right)$
to decide whether to actually accept the moves or not). Also, instead
of starting from a random configuration, we can use the BP marginals
in the absence of any distance constraint, and clip them to determine
a good starting point.

\begin{figure}
\includegraphics[width=1\textwidth]{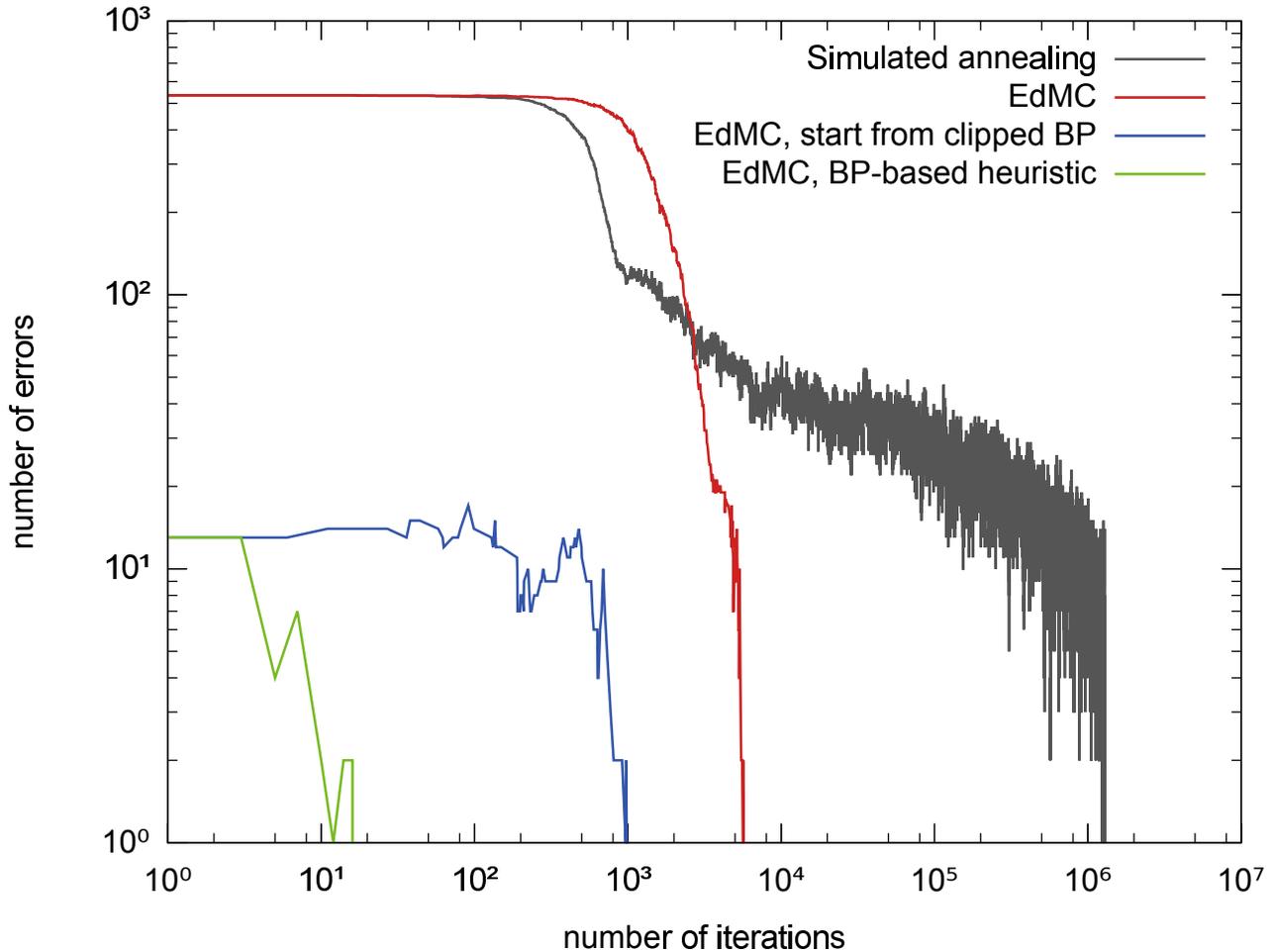}\protect\caption{\label{fig:EdMC}Comparison between different Monte Carlo-based solver
algorithms for one sample with $N=501$, $\alpha=1.2$, $L=4$ and
$f=0.1$. The curves show in log-log scale the number of errors of
the system as a function of the number of iterations (note that while
the number of errors is used as the energy throughout the rest of
the paper, none of the algorithms shown here uses it as its objective
function). The curves shown are labeled in worst to best order: simulated
annealing on $E_{\Delta}$ (gray curve, see eq.~\eqref{eq:en_neg_stab},
more than $10^{6}$ iterations required to find a solution); EdMC
starting from random initial condition with zero-temperature dynamics
(red curve, less than $10^{4}$ iterations); EdMC using BP marginals
as initial condition with zero-temperature dynamics (blue curve, less
than $10^{3}$ iterations); EdMC using BP marginals both as initial
condition and to propose the Monte Carlo moves (green curve, less
than $10^{2}$ iterations). The local-entropy landscape is clearly
much smoother than the energy landscape (even when using the energy
$E_{\Delta}$).}
\end{figure}

Fig.~\ref{fig:EdMC} shows the results of a test on one sample for
$N=501$, $\alpha=1.2$, $L=4$, $f=0.1$. Although the search space
is considerably larger, the behavior of the algorithm is very similar
to what was observed in~\cite{baldassi_local_2016} for the binary,
balanced and unbiased case: while EdMC reaches $0$ errors in a few
iterations, standard SA plateaus and only eventually finds a solution,
in several orders of magnitude more iterations (as is typical for
these glassy systems, the time during which SA is trapped in a plateau
increases exponentially with $N$). The heuristic enhancements further
improve EdMC performance.

More specifically in this test all the variants of the EdMC were run
at $y=\infty$; when the initial configuration was chosen at random,
we started with external fields of low intensity $\gamma=0.5$ and
progressively increased it by $\Delta\gamma=1.0$ after each greedy
optimization procedure, thus avoiding inconsistencies in the messages
and guaranteeing the convergence of BP even in the early stages, when
the reference configuration $\tilde{W}$ is very far away from any
solution. When starting from the clipped BP marginals the fields could
be set directly at $\gamma=3.5$.

On the other hand in the SA we observed that the chosen $\alpha$
was large enough to trap the standard Monte Carlo even with very slow
cooling rates, so we had to resort to a different definition of the
energy function 
\begin{equation}
E_{\Delta}\left(\tilde{W}\right)=\sum_{\mu}\left(-\left(2\sigma^{\mu}-1\right)\left(\sum_{i}\tilde{W}_{i}\xi_{i}^{\mu}-\theta N\right)\right)_{+}\label{eq:en_neg_stab}
\end{equation}
 where $\left(x\right)_{+}=x$ if $x>0$, $0$ otherwise; i.e.~this
energy function measures the negative of the sum of the so-called
stabilities. The annealing scheme was carried out adopting a cooling
rate of $r_{y}=1.005$, which is multiplied to $y$ after every $100$
accepted moves, starting from an inverse temperature of $y=1.0$.

In both the EdMC and the SA the firing threshold $\theta$ was set
to its optimal value, which was determined analytically via replica
calculations.

\section{Conclusions}

In this work, we extended a large deviation analysis of the solution
space of a single layer neural network from the purely binary and
balanced case \cite{baldassi_subdominant_2015} to the general discrete
case. Despite some technical challenges in solving the equations,
the results clearly indicate that the general qualitative picture
is unchanged with respect to the binary case. In particular, for all
values of the parameters and regardless of the number of synaptic
states, we observe the existence of two distinct phases: one in which
most solutions are isolated and hard to find, but there exists a dense
and accessible cluster of solutions whose presence can be directly
or indirectly exploited by heuristic solvers (e.g.~by the EdMC algorithm);
one in which this dense cluster has broken apart. The transition point
$\alpha_{U}$ between these two phases was greater than $0.9\alpha_{c}$
in all our tested settings, i.e.~it is fairly close to the maximal
theoretical capacity. Both $\alpha_{c}$ and $\alpha_{U}$ grow with
the number of synaptic states; however, the increase becomes rapidly
very slow after the first few states: if there is a cost (metabolic
or hardware) associated to adding more states to the synapses, this
analysis suggests that the overall benefit of doing so would rapidly
vanish. In other words, synapses may only need few bits of precision,
both in the sense that efficient learning is still possible (contrary
to what previous analyses suggested) and in the sense that, increasing
the precision, the marginal advantage in terms of capacity decreases
quite rapidly.

Our main drive for performing this analysis was to make the model
more biologically plausible with respect to the binary case, while
still keeping it simple enough so that the theoretical analysis can
be performed (albeit with great difficulty, for computational and
numerical reasons). Indeed, as we already mentioned, our model neurons
are very crude simplifications of biological neurons; also, using
uncorrelated inputs and outputs is hardly realistic, or at the very
least there certainly are settings in which we would rather consider
some kind of correlations. Despite these shortcomings, we believe
that this analysis, together with the previous one for the binary
case, bears a rather clear general message, namely that the qualitative
picture is the same regardless of the finer detail. In particular,
this picture was not affected in our analysis by any of the parameters
(number of synaptic states, sparsity of the patterns). Also, we had
already shown for the binary case that numerical tests performed on
a handwritten-digit image-recognition benchmark indicate that even
when using more ``natural'' (highly correlated and structured) patterns
the heuristic learning algorithms invariably end up in a dense region
of solutions such as those described by the theoretical analysis of
the uncorrelated-inputs case.

Therefore, despite the inevitable shortcomings of the model, this
analysis provides a plausible picture in which to frame the study
of the synaptic precision of biological neurons in relation to their
computational and representational power. In a nutshell, it suggests
that low precision synapses are convenient for concrete implementations
because the solution space has regions that can be exploited for learning
efficiently, consistently with experimental biological results. Indeed,
the learning mechanism must be different from what is usually employed
in machine learning applications (stochastic gradient descent), but
simple effective algorithms exist thanks to the peculiar structure
of the solution space, with ample room for discretion in implementation
details. This is clearly exemplified by the effectiveness of the Entropy-driven
Monte Carlo technique that we introduced in~\cite{baldassi_local_2016}
and that we extended here to the more general case. Establishing the
presence of this geometrical picture in the learning of discrete deep
forward networks and recurrent neural networks looks like a promising
direction for future investigations. 
\begin{acknowledgments}
C.B., C.L. and R.Z. acknowledge the European Research Council for
grant n\textdegree ~267915.
\end{acknowledgments}

\appendix

\section{Franz-Parisi potential\label{sec:AppendixA-FP}}

In order to describe the geometrical properties of the solution space
of the generalized perceptron, it is possible to carry out a mean-field
analysis based on the computation of the Franz-Parisi potential. This
method is conceptually divided in two stages: first we select a reference
configuration $\tilde{W}$ from the equilibrium Boltzmann measure
at a certain inverse temperature $\beta^{\prime}$, then we evaluate
the free energy of a coupled model where the configurations $\{W\}$,
at inverse temperature $\beta$, are constrained to be exactly at
a distance $D$ from the reference point:

\begin{equation}
\mathcal{S}_{FP}\left(\beta^{\prime},\beta,D\right)=\frac{1}{N}\left\langle \frac{1}{Z\left(\beta^{\prime}\right)}\sum_{\left\{ \tilde{W}\right\} }e^{-\beta^{\prime}E\left(\tilde{W}\right)}\log\left(\sum_{\left\{ W\right\} }e^{-\beta E\left(W\right)}\delta\left(d\left(W,\tilde{W}\right)-D\right)\right)\right\rangle _{\{\xi,\sigma\}}\label{eq:franz-parisi_temperatures-1-1}
\end{equation}

Since we are interested in the constraint satisfaction problem, in
our case both temperatures are set to zero ($\beta,\beta^{\prime}\to\infty$).
It is important to notice that the sampling of $\tilde{W}$ is not
affected by the coupling, so it is extracted at random from a flat
distribution over all possible solutions, and represents the \emph{typical}
case (i.e.~numerically dominant in this measure).

The Franz-Parisi potential can thus be interpreted as a typical \emph{local
entropy} density:
\begin{equation}
\mathcal{S}_{FP}\left(D\right)=\frac{1}{N}\left\langle \left\langle \log\sum_{\left\{ W\right\} }\mathbb{X}_{\xi,\sigma}\left(W\right)\delta\left(d\left(W,\tilde{W}\right)-D\right)\right\rangle _{\tilde{W}}\right\rangle _{\left\{ \xi,\sigma\right\} }\label{eq:typical_local_entropy_meanWt}
\end{equation}

with the definition of the indicator function $\mathbb{X}_{\xi,\sigma}\left(W\right)$
of equation \ref{eq:indicator} and the averaging $\left\langle \cdot\right\rangle _{\tilde{W}}$
is performed over the flat measure on all solutions to the problem.

It is possible to introduce a robustness parameter $K$ to stabilize
the learned patterns (at the order $O\left(\sqrt{N}\right)$), so
that each association is considered learned only if the output of
the device is correct and the modulus of the activation $\left|\sum_{i=1}^{N}\frac{W_{i}\xi_{i}^{\mu}}{\sqrt{N}}-\theta\sqrt{N}\right|$
is above this threshold. The indicator function can be then redefined
as: 

\begin{equation}
\mathbb{X}_{\xi,\sigma}\left(W,K\right)=\prod_{\mu}\Theta\left(s^{\mu}\left(\sum_{i}\frac{W_{i}\xi_{i}^{\mu}}{\sqrt{N}}-\theta\sqrt{N}\right)-K\right)\label{eq:indicator_withK}
\end{equation}
where we omitted the indication of the ranges $i\in\left\{ 1,\dots,N\right\} $
and $\mu\in\left\{ 1,\dots,\alpha N\right\} $ for simplicity of notation,
and we defined $s^{\mu}=2\sigma^{\mu}-1$ in order to convert between
the device output $\sigma^{\mu}\in\left\{ 0,1\right\} $ and a more
convenient representation $s^{\mu}\in\left\{ -1,+1\right\} $. Note
that with this definition the average over the output $\sigma^{\mu}$
for any function $g\left(s^{\mu}\right)$ is defined as:
\begin{equation}
\left\langle g\left(s^{\mu}\right)\right\rangle _{s^{\mu}}=f^{\prime}g\left(1\right)+\left(1-f^{\prime}\right)g\left(-1\right)\label{eq:output_average}
\end{equation}
i.e.~we use the parameter $f^{\prime}$ to denote the output coding
rate, which in principle can be distinguished from the input coding
rate $f$.

In order to perform the average over the measure of solutions $\tilde{W}$
and over the quenched disorder, we employ the replica trick: we introduce
$\tilde{n}-1$ non interacting copies $\tilde{W}^{c}$ of the reference
solution, and leave out the index $\tilde{W}^{c=1}$ for the replica
appearing in the distance constraint. Furthermore we denote the $n$
replicas of the coupled solutions $W^{a}$. Throughout this section,
we will use the indices $a,b\in\left\{ 1,\dots,n\right\} $ for the
replicated $W$ and $c,d\in\left\{ 1,\dots,\tilde{n}\right\} $ for
the replicated $\tilde{W}$, and we will omit the specification of
the indices ranges in sums and products, for notational simplicity.
In the end $\tilde{n}$ and $n$ will be sent to zero:

\begin{eqnarray}
\mathcal{S}_{FP}\left(D\right) & = & \frac{1}{N}\lim_{n,\tilde{n}\rightarrow0}\frac{\partial}{\partial n}\left<\int\prod_{i,c}d\mu\left(\tilde{W}_{i}^{c}\right)\int\prod_{i,a}d\mu\left(W_{i}^{a}\right)\prod_{c}\mathbb{X}_{\xi,\sigma}\left(\tilde{W}^{c},K\right)\prod_{a}\mathbb{X}_{\xi,\sigma}\left(W^{a},K\right)\right.\label{eq:free_en_FP}\\
 &  & \qquad\qquad\qquad\left.\times\prod_{a}\delta\left(\frac{1}{2}\sum_{i}\left(W_{i}^{a}-\tilde{W}_{i}^{1}\right)^{2}-2DN\right)\right>_{\xi,\sigma}\nonumber \\
 & \equiv & \frac{1}{N}\lim_{n\rightarrow0}\frac{\partial}{\partial n}\Omega_{FP}^{n}\left(D\right)\nonumber 
\end{eqnarray}
where we used the definition of eq.~\eqref{eq:distance} for the
distance function $d\left(\cdot,\cdot\right)$ and introduced the
measure over the possible values of the weights: 
\begin{equation}
d\mu\left(W\right)=\sum_{l\in\mathcal{L}}\delta\left(W-l\right)\label{eq:weights_measure}
\end{equation}
(In our experiments, we always used $\mathcal{L}=\left\{ 0,1,\dots,L\right\} $,
but the derivation is general.) In the last line of eq.~\eqref{eq:free_en_FP}
we also defined the replicated volume $\Omega_{FP}^{n}\left(D\right)$.

As a first step we can introduce some auxiliary variables to substitute
the arguments of the indicator functions: 

\begin{eqnarray}
 &  & \prod_{c}\mathbb{X}_{\xi,\sigma}\left(\tilde{W}^{c},K\right)\prod_{a}\mathbb{X}_{\xi,\sigma}\left(W^{a},K\right)=\\
 &  & =\int\prod_{\mu,a}\frac{d\lambda_{\mu}^{a}d\hat{\lambda}_{\mu}^{a}}{2\pi}\int\prod_{\mu,c}\frac{d\tilde{\lambda}_{\mu}^{c}d\hat{\tilde{\lambda}}_{\mu}^{c}}{2\pi}\left\langle \prod_{\mu,a}\Theta\left(\sigma^{\mu}\lambda_{\mu}^{a}-K\right)\prod_{\mu,c}\Theta\left(\sigma^{\mu}\tilde{\lambda}_{\mu}^{c}-K\right)\right\rangle _{\sigma}\prod_{\mu,a}e^{i\lambda_{\mu}^{a}\hat{\lambda}_{\mu}^{a}}\prod_{\mu,c}e^{i\tilde{\lambda}_{\mu}^{c}\hat{\tilde{\lambda}}_{\mu}^{c}}\times\nonumber \\
 &  & \quad\times\prod_{\mu,i}\left(e^{i\theta\sqrt{N}\sum_{ac}\hat{\lambda}_{\mu}^{ca}}\left\langle \exp\left(-\frac{i}{\sqrt{N}}\left(\sum_{a}\hat{\lambda}_{\mu}^{a}W_{i}^{a}+\sum_{c}\hat{\tilde{\lambda}}_{\mu}^{c}\tilde{W}_{i}^{c}\right)\xi_{i}^{\mu}\right)\right\rangle _{\xi}\right)\nonumber 
\end{eqnarray}

We can now perform the average over the pattern distribution $\left\langle \cdot\right\rangle _{\xi}=\int\prod_{i,\mu}\left(P\left(\xi_{i}^{\mu}\right)d\xi_{i}^{\mu}\right)$:

\begin{eqnarray}
 &  & \prod_{\mu,i}\left\langle \exp\left(-\frac{i}{\sqrt{N}}\left(\sum_{a}\hat{\lambda}_{\mu}^{a}W_{i}^{a}+\sum_{c}\hat{\tilde{\lambda}}_{\mu}^{c}\tilde{W}_{i}^{c}\right)\xi_{i}^{\mu}\right)\right\rangle _{\xi}=\label{eq:disorder_av}\\
 &  & =\prod_{\mu}\exp\sum_{i}\log\left(1-\frac{i}{\sqrt{N}}\left(\sum_{a}\hat{\lambda}_{\mu}^{a}W_{i}^{a}+\sum_{c}\hat{\tilde{\lambda}}_{\mu}^{c}\tilde{W}_{i}^{c}\right)\overline{\xi}+\right.\nonumber \\
 &  & \qquad\left.-\frac{1}{2N}\left(\left(\sum_{a}\hat{\lambda}_{\mu}^{a}W_{i}^{a}\right)^{2}+\left(\sum_{c}\hat{\tilde{\lambda}}_{\mu}^{c}\tilde{W}_{i}^{c}\right)^{2}+2\sum_{ac}\hat{\lambda}_{\mu}^{a}\hat{\tilde{\lambda}}_{\mu}^{c}W_{i}^{a}\tilde{W}_{i}^{c}\right)\overline{\xi^{2}}\right)\nonumber \\
 &  & =\prod_{\mu}\exp\left(-i\overline{\xi}\sqrt{N}\left(\sum_{a}\hat{\lambda}_{\mu}^{a}\sum_{i}\frac{W_{i}^{a}}{N}+\sum_{c}\hat{\tilde{\lambda}}_{\mu}^{c}\sum_{i}\frac{W_{i}^{c}}{N}\right)+\right.\nonumber \\
 &  & \qquad\left.-\frac{\sigma_{\xi}^{2}}{2}\left(\sum_{ab}\hat{\lambda}_{\mu}^{a}\hat{\lambda}_{\mu}^{b}\sum_{i}\frac{W_{i}^{a}W_{i}^{b}}{N}+\sum_{cd}\hat{\lambda}_{\mu}^{c}\hat{\lambda}_{\mu}^{d}\sum_{i}\frac{\tilde{W}_{i}^{c}\tilde{W}_{i}^{d}}{N}+2\sum_{ac}\hat{\lambda}_{\mu}^{a}\hat{\tilde{\lambda}}_{\mu}^{c}\sum_{i}\frac{W_{i}^{a}\tilde{W}_{i}^{c}}{N}\right)\right)\nonumber 
\end{eqnarray}

where $\overline{\xi}$ indicates the average of the inputs and $\sigma_{\xi}^{2}$
is their variance.

All the overlaps (such as $\frac{1}{N}\sum_{i}W_{i}^{a}W_{i}^{b}$)
can now be replaced with order parameters via Dirac $\delta$ distributions.
In the case of the generalized perceptron we also need to introduce
two specific parameters for the $L^{1}$-norm and for the $L^{2}$-norm.

Maximum capacity with biased patterns can be achieved if the mean
value of the synaptic weights $\overline{W}$ is on the threshold
given by the ratio:

\begin{eqnarray}
\overline{W} & = & \frac{\theta}{f}
\end{eqnarray}

and because of the unbalanced distribution of the outputs we also
need to introduce an $O\left(\frac{1}{\sqrt{N}}\right)$ correction
controlled by the order parameter $M$:

\begin{eqnarray}
\sum_{i}\frac{W_{i}}{N} & = & \overline{W}+\frac{M}{\sqrt{N}}
\end{eqnarray}

We can define the following:
\begin{itemize}
\item $\sum_{i}\frac{\left(W_{i}^{a}\right)^{2}}{N}=Q^{a}$, $\sum_{i}\frac{\left(\tilde{W}_{i}^{c}\right)^{2}}{N}=\tilde{Q^{c}}$
\item $\sum_{i}\frac{W_{i}^{a}}{N}=\overline{W}+\frac{M^{a}}{\sqrt{N}}$,
$\sum_{i}\frac{\tilde{W}_{i}^{c}}{N}=\overline{\tilde{W}}+\frac{\tilde{M}^{c}}{\sqrt{N}}$
\item $\sum_{i}\frac{W_{i}^{a}W_{i}^{b}}{N}=q^{ab}$, $\sum_{i}\frac{\tilde{W}_{i}^{c}\tilde{W}_{i}^{d}}{N}=\tilde{q}^{cd}$,
$\sum_{i}\frac{W_{i}^{a}\tilde{W}_{i}^{c}}{N}=S^{ca}$
\end{itemize}
After these substitutions in the expression of the replicated volume
$\Omega^{n}\left(D\right)$, we use the integral representation of
the Dirac $\delta$ distributions, introducing the required conjugate
parameters, and rearrange the integrals so that it becomes possible
to factorize over the $\mu$ and $i$ indices:

\begin{eqnarray}
\Omega_{FP}^{n}\left(D\right) & = & \lim_{\tilde{n}\rightarrow0}\int\prod_{c>d}\frac{d\tilde{q}^{cd}d\hat{\tilde{q}}^{cd}}{\left(2\pi/N\right)}\int\prod_{a>b}\frac{dq^{ab}d\hat{q}^{ab}}{\left(2\pi/N\right)}\int\prod_{c}\frac{d\tilde{Q}^{c}d\hat{\tilde{Q}}^{c}}{\left(2\pi/N\right)}\int\prod_{a}\frac{dQ^{a}d\hat{Q}^{a}}{\left(2\pi/N\right)}\\
 &  & \qquad\int\prod_{c}\frac{d\tilde{M}^{c}d\hat{\tilde{M}}^{c}}{\left(2\pi/\sqrt{N}\right)}\int\prod_{a}\frac{dM^{a}d\hat{M}^{a}}{\left(2\pi/\sqrt{N}\right)}\int\prod_{ca}\frac{dS^{ca}d\hat{S}^{ca}}{\left(2\pi/N\right)}\nonumber \\
 &  & \qquad\int\prod_{a}\frac{d\hat{D}^{a}}{2\pi}G_{1}\,\left(G_{S}\right)^{N}\,\left(G_{E}\right)^{\alpha N}\nonumber 
\end{eqnarray}

where we have singled out a first term $G_{1}$ and the so-called
entropic and energetic contributions $G_{S}$, $G_{E}$:

\begin{eqnarray}
G_{1} & = & \exp\left(-N\left(\sum_{c>d}\hat{\tilde{q}}^{cd}\tilde{q}^{cd}+\sum_{a>b}\hat{q}^{ab}q^{ab}+\sum_{c}\hat{\tilde{Q}}^{c}\tilde{Q}^{c}+\sum_{a}\hat{Q}^{a}Q^{a}+\sum_{c}\hat{\tilde{M}}^{c}\left(\frac{\tilde{M}^{c}}{\sqrt{N}}+\overline{\tilde{W}}\right)+\right.\right.\\
 &  & \left.\left.+\sum_{a}\hat{M}^{a}\left(\frac{M^{a}}{\sqrt{N}}+\overline{W}\right)+\sum_{ca}\hat{S}^{ca}S^{ca}+\sum_{a}\hat{D}^{a}\left(\frac{1}{2}Q^{a}+\frac{1}{2}\tilde{Q}^{1}-S^{1a}-2D\right)\right)\right)\nonumber \\
G_{S} & = & \int\prod_{c}d\mu\left(\tilde{W}^{c}\right)\int\prod_{a}d\mu\left(W^{a}\right)\exp\left(\sum_{c>d}\hat{\tilde{q}}^{cd}\tilde{W}^{c}\tilde{W}^{d}+\sum_{a>b}\hat{q}^{ab}W^{a}W^{b}+\right.\\
 &  & \left.+\sum_{c}\hat{\tilde{Q}}^{c}\left(\tilde{W}^{c}\right)^{2}+\sum_{a}\hat{Q}^{a}\left(W^{a}\right)^{2}+\sum_{c}\hat{\tilde{M}}^{c}\tilde{W}^{c}+\sum_{a}\hat{M}^{a}W^{a}+\sum_{ca}\hat{S}^{ca}W^{a}\tilde{W}^{c}\right)\nonumber \\
G_{E} & = & \int\prod_{a}\frac{d\lambda^{a}d\hat{\lambda}^{a}}{2\pi}\int\prod_{c}\frac{d\tilde{\lambda}^{c}d\hat{\tilde{\lambda}}^{c}}{2\pi}\left\langle \prod_{a}\Theta\left(s\lambda^{a}-K\right)\prod_{c}\Theta\left(s\tilde{\lambda}^{c}-K\right)\right\rangle _{s}\times\\
 &  & \times\exp\left(i\left(\sum_{a}\lambda^{a}\hat{\lambda}^{a}+\sum_{c}\tilde{\lambda}^{c}\hat{\tilde{\lambda}}^{c}-\overline{\xi}\sum_{a}\hat{\lambda}^{a}M^{a}-\overline{\xi}\sum_{c}\hat{\tilde{\lambda}}^{c}\tilde{M}^{c}\right)+\right.\nonumber \\
 &  & \left.\quad-\frac{1}{2}\sigma_{\xi}^{2}\sum_{a}\left(\hat{\lambda}^{a}\right)^{2}Q^{a}-\frac{1}{2}\sigma_{\xi}^{2}\sum_{c}\left(\hat{\tilde{\lambda}}^{c}\right)^{2}\tilde{Q}^{c}-\frac{1}{2}\sigma_{\xi}^{2}\sum_{(a,b)}\hat{\lambda}^{a}\hat{\lambda}^{b}q^{ab}-\frac{1}{2}\sigma_{\xi}^{2}\sum_{(c,d)}\hat{\tilde{\lambda}}^{c}\hat{\tilde{\lambda}}^{d}\tilde{q}^{cd}-\sigma_{\xi}^{2}\sum_{ac}\hat{\lambda}^{a}\hat{\tilde{\lambda}}^{c}S^{ac}\right)\nonumber 
\end{eqnarray}

Note that we dropped the indices $i$ and $\mu$ from all quantities
since we have rearranged the terms and factorized the contributions;
in particular, note that the indices were dropped from the weights
$W^{a}$, $\tilde{W}^{c}$ and the output $s$.

\subsection{Replica Symmetric Ansatz}

To proceed with the calculations we now have to make an assumption
on the structure of the replicated order parameters, the simplest
possible one being the symmetric Ansatz, where one can drop all the
dependencies on the replica indices. We only have to make a distinction
between the overlaps $S$ and $\tilde{S}$, since the first one enters
also in the expression of the constraint on the distance $d\left(W,\tilde{W}\right)$:
\begin{itemize}
\item $S^{ca}=S$ for $c=1$, $S^{ca}=\tilde{S}$ for $c\neq1$
\item $Q^{a}=Q$, $\tilde{Q}^{c}=\tilde{Q}$, $M^{a}=M$, $\tilde{M}^{c}=\tilde{M}$,
$q^{ab}=q$, $\tilde{q}^{cd}=\tilde{q}$, $\hat{D}^{ca}=\hat{D}$.
\end{itemize}
The first term $G_{1}$ of the expression of the volume can now be
simplified, and the $\tilde{n}\to0$ can be taken, obtaining:

\begin{eqnarray}
 & G_{1} & =\lim_{\tilde{n}\rightarrow0}\exp\left(-N\left(\frac{\tilde{n}\left(\tilde{n}-1\right)}{2}\hat{\tilde{q}}\tilde{q}+\frac{n\left(n-1\right)}{2}\hat{q}q+\tilde{n}\hat{\tilde{Q}}\tilde{Q}+n\hat{Q}Q+\right.\right.\\
 &  & \left.\left.\qquad+\tilde{n}\hat{\tilde{M}}\overline{\tilde{W}}+n\hat{M}\overline{W}+n\hat{S}S-\left(1-\tilde{n}\right)n\hat{\tilde{S}}\tilde{S}+n\hat{D}\left(\frac{1}{2}Q+\frac{1}{2}\tilde{Q}-S-2D\right)\right)\right)\nonumber \\
 &  & =\exp\left(-Nn\left(-\frac{1}{2}\hat{q}q+\hat{Q}Q+\hat{M}\overline{W}+\hat{S}S-\hat{\tilde{S}}\tilde{S}+\hat{D}\left(\frac{1}{2}Q+\frac{1}{2}\tilde{Q}-S-2D\right)\right)\right)\nonumber 
\end{eqnarray}

After the substitution of the RS Ansatz, the entropic term reads:

\begin{eqnarray}
G_{S} & = & \int\prod_{c}d\mu\left(\tilde{W}^{c}\right)\int\prod_{a}d\mu\left(W^{a}\right)\exp\left(\left(\hat{\tilde{Q}}-\frac{1}{2}\hat{\tilde{q}}\right)\sum_{c}\left(\tilde{W}^{c}\right)^{2}+\left(\hat{Q}-\frac{1}{2}\hat{q}\right)\sum_{a}\left(W^{a}\right)^{2}+\right.\\
 &  & +\frac{1}{2}\hat{\tilde{q}}\left(\sum_{c}\tilde{W}^{c}\right)^{2}+\frac{1}{2}\hat{q}\left(\sum_{a}W^{a}\right)^{2}+\hat{\tilde{M}}\sum_{c}\tilde{W}^{c}+\hat{M}\sum_{a}W^{a}+\nonumber \\
 &  & \left.+\left(\hat{S}-\hat{\tilde{S}}\right)\sum_{a}W^{a}\tilde{W}^{1}+\hat{\tilde{S}}\sum_{a}W^{a}\sum_{c}\tilde{W}^{c}\right)\nonumber 
\end{eqnarray}

Now, in order to be able to factorize over the replica index $c$,
we need to write:

\[
\hat{\tilde{S}}\sum_{a}W^{a}\sum_{c}\tilde{W}^{c}=\frac{1}{2}\hat{\tilde{S}}\left(\sum_{a}W^{a}+\sum_{c}\tilde{W}^{c}\right)^{2}-\frac{1}{2}\hat{\tilde{S}}\left(\sum_{a}W^{a}\right)^{2}-\frac{1}{2}\hat{\tilde{S}}\left(\sum_{c}\tilde{W}^{c}\right)^{2}
\]

and then we perform some Hubbard-Stratonovich transformations, introducing
the variables $x$, $z$ and $\tilde{z}$.

Using the usual notation $\int\mathcal{D}z=\int_{-\infty}^{+\infty}\frac{dz}{\sqrt{2\pi}}e^{-\frac{z^{2}}{2}}$
for Gaussian integrals, we get:

\begin{eqnarray}
G_{S} & = & \int\mathcal{D}x\int\mathcal{D}z\int\mathcal{D}\tilde{z}\\
 &  & \int\prod_{c}d\mu\left(\tilde{W}^{c}\right)\exp\left(\left(\hat{\tilde{Q}}-\frac{1}{2}\hat{\tilde{q}}\right)\sum_{c}\left(\tilde{W}^{c}\right)^{2}+\left(\tilde{z}\sqrt{\hat{\tilde{q}}-\hat{\tilde{S}}}+x\sqrt{\hat{\tilde{S}}}+\hat{\tilde{M}}\right)\sum_{c}\tilde{W}^{c}\right)\times\nonumber \\
 &  & \times\int\prod_{a}d\mu\left(W^{a}\right)\exp\left(\left(\hat{Q}-\frac{1}{2}\hat{q}\right)\sum_{a}\left(W^{a}\right)^{2}+\left(z\sqrt{\hat{q}-\hat{\tilde{S}}}+x\sqrt{\hat{\tilde{S}}}+\hat{M}+\left(\hat{S}-\hat{\tilde{S}}\right)\tilde{W}^{1}\right)\sum_{a}W^{a}\right)\nonumber 
\end{eqnarray}

Since the expression is now factorized, with the definitions:

\begin{eqnarray}
\tilde{A}\left(\tilde{W},\tilde{z},x\right) & = & \left(\hat{\tilde{Q}}-\frac{1}{2}\hat{\tilde{q}}\right)\tilde{W}^{2}+\left(\tilde{z}\sqrt{\hat{\tilde{q}}-\hat{\tilde{S}}}+x\sqrt{\hat{\tilde{S}}}+\tilde{M}\right)\tilde{W}\\
A\left(W,\tilde{W},z,x\right) & = & \left(\hat{Q}-\frac{1}{2}\hat{q}\right)W^{2}+\left(z\sqrt{\hat{q}-\hat{\tilde{S}}}+x\sqrt{\hat{\tilde{S}}}+\hat{M}+\Delta\hat{S}\tilde{W}\right)W\\
\Delta\hat{S} & = & \hat{S}-\hat{\tilde{S}}
\end{eqnarray}

we can first take the limit $\tilde{n}\to0$, restoring the presence
of the denominator:

\begin{eqnarray}
G_{S} & = & \int\mathcal{D}x\int\mathcal{D}z\int\mathcal{D}\tilde{z}\frac{\int d\mu\left(\tilde{W}\right)\exp\left(\tilde{A}\left(\tilde{W},\tilde{z},x\right)\right)\int\prod_{a}d\mu\left(W^{a}\right)\prod_{a}\exp\left(A^{a}\left(W^{a},\tilde{W},z,x\right)\right)}{\int d\mu\left(\tilde{W}\right)\exp\left(\tilde{A}\left(\tilde{W},\tilde{z},x\right)\right)}
\end{eqnarray}

And then, in the limit $n\to0$, we can write:

\begin{eqnarray}
\mathcal{G}_{S}=\frac{1}{n}\log G_{S} & = & \int\mathcal{D}x\int\mathcal{D}z\int\mathcal{D}\tilde{z}\frac{\int d\mu\left(\tilde{W}\right)\exp\left(\tilde{A}\left(\tilde{W},\tilde{z},x\right)\right)\log\left(\int d\mu\left(W\right)\exp\left(A\left(W,\tilde{W},z,x\right)\right)\right)}{\int d\mu\left(\tilde{W}\right)\exp\left(\tilde{A}\left(\tilde{W},\tilde{z},x\right)\right)}
\end{eqnarray}

Then we perform two rotations between the integration variables ($\tilde{z}$,
$x$) and ($z$, $x$), in order to compute analytically the $\int\mathcal{D}x$
integral, obtaining:

\begin{eqnarray}
 &  & \mathcal{G}_{S}=\\
 &  & \int\negthickspace\mathcal{D}z\int\negthickspace\mathcal{D}\tilde{z}\frac{\sum_{\tilde{l}}\exp\left(\left(\hat{\tilde{Q}}-\frac{1}{2}\hat{\tilde{q}}\right)\tilde{l}^{2}+\left(\hat{\tilde{M}}+\tilde{z}\sqrt{\hat{q}}\right)\tilde{l}\right)\log\left(\sum_{l}\exp\left(\left(\hat{Q}-\frac{1}{2}\hat{q}\right)l^{2}+\left(\hat{M}+z\sqrt{\frac{\hat{q}\hat{\tilde{q}}-\hat{\tilde{S}}^{2}}{\hat{\tilde{q}}}}+\tilde{z}\frac{\hat{\tilde{S}}}{\sqrt{\hat{\tilde{q}}}}+\Delta\hat{S}\,\,\tilde{l}\right)l\right)\right)}{\sum_{\tilde{l}}\exp\left(\left(\hat{\tilde{Q}}-\frac{1}{2}\hat{\tilde{q}}\right)\tilde{l}^{2}+\left(\hat{\tilde{M}}+\tilde{z}\sqrt{\hat{q}}\right)\tilde{l}\right)}\nonumber 
\end{eqnarray}

We proceed in a similar way with the computation of the energetic
term:

\begin{eqnarray}
G_{E} & = & \int\mathcal{D}x\int\prod_{a}\frac{d\lambda^{a}d\hat{\lambda}^{a}}{2\pi}\int\prod_{c}\frac{d\tilde{\lambda}^{c}d\hat{\tilde{\lambda}}^{c}}{2\pi}\left\langle \prod_{a}\Theta\left(s\lambda^{a}-K\right)\prod_{c}\Theta\left(s\tilde{\lambda}^{c}-K\right)\right\rangle _{s}\times\\
 &  & \times\exp\left(i\sum_{a}\hat{\lambda}^{a}\left(\lambda^{a}-\overline{\xi}M-x\sqrt{\sigma_{\xi}^{2}\tilde{S}}\right)+i\sum_{c}\hat{\tilde{\lambda}}^{c}\left(\tilde{\lambda}^{c}-\overline{\xi}\tilde{M}-x\sqrt{\sigma_{\xi}^{2}\tilde{S}}\right)-\frac{1}{2}\sigma_{\xi}^{2}\left(Q-q\right)\sum_{a}\left(\hat{\lambda}^{a}\right)^{2}+\right.\nonumber \\
 &  & \left.-\frac{1}{2}\sigma_{\xi}^{2}\left(\tilde{Q}-\tilde{q}\right)\sum_{c}\left(\hat{\tilde{\lambda}}^{c}\right)^{2}-\frac{1}{2}\sigma_{\xi}^{2}\left(q-\tilde{S}\right)\left(\sum_{a}\hat{\lambda}^{a}\right)^{2}-\frac{1}{2}\sigma_{\xi}^{2}\left(\tilde{q}-\tilde{S}\right)\left(\sum_{c}\hat{\tilde{\lambda}}^{c}\right)^{2}-\sigma_{\xi}^{2}\left(S-\tilde{S}\right)\hat{\tilde{\lambda}}^{1}\sum_{a}\hat{\lambda}^{a}\right)\nonumber 
\end{eqnarray}

We define:

\begin{eqnarray}
\tilde{B}\left(\tilde{\lambda},\tilde{z},x\right) & = & -\frac{1}{2}\sigma_{\xi}^{2}\left(\tilde{Q}-\tilde{q}\right)\hat{\tilde{\lambda}}^{2}+i\left(\tilde{\lambda}-\overline{\xi}\tilde{M}-x\sqrt{\sigma_{\xi}^{2}\tilde{S}}-\tilde{z}\sqrt{\sigma_{\xi}^{2}\left(\tilde{q}-\tilde{S}\right)}\right)\hat{\tilde{\lambda}}\\
B\left(\lambda,\tilde{\lambda},z,x\right) & = & -\frac{1}{2}\sigma_{\xi}^{2}\left(Q-q\right)\hat{\lambda}^{2}+i\left(\lambda-\overline{\xi}M-x\sqrt{\sigma_{\xi}^{2}\tilde{S}}-z\sqrt{\sigma_{\xi}^{2}\left(q-\tilde{S}\right)}+i\sigma_{\xi}^{2}\left(S-\tilde{S}\right)\hat{\tilde{\lambda}}\right)\hat{\lambda}
\end{eqnarray}

and in the $n\to0$ limit we find:

\begin{eqnarray}
 &  & \mathcal{G}_{E}=\frac{1}{n}\log G_{E}=\\
 &  & \int\mathcal{D}x\int\mathcal{D}z\int\mathcal{D}\tilde{z}\left\langle \frac{\int\frac{d\tilde{\lambda}d\hat{\tilde{\lambda}}}{2\pi}\Theta\left(s\tilde{\lambda}-K\right)\exp\left(\tilde{B}\left(\tilde{\lambda},\tilde{z},x\right)\right)\log\left(\int\frac{d\lambda d\hat{\lambda}}{2\pi}\Theta\left(s\lambda-K\right)\exp\left(B\left(\lambda,\tilde{\lambda},z,x\right)\right)\right)}{\int\frac{d\tilde{\lambda}d\hat{\tilde{\lambda}}}{2\pi}\Theta\left(s\tilde{\lambda}-K\right)\exp\left(\tilde{B}\left(\tilde{\lambda},\tilde{z},x\right)\right)}\right\rangle _{s}\nonumber 
\end{eqnarray}

We leave the output average written implicitly (see eq.~\eqref{eq:output_average})
for simplicity. We can evaluate the $\hat{\lambda}$ and $\lambda$
integrals introducing the normalized integral function:

\begin{eqnarray}
H\left(x\right) & =\int_{x}^{\infty}\mathcal{D}x= & \frac{1}{2}\erfc\left(\frac{x}{\sqrt{2}}\right)\label{eq:h_function}
\end{eqnarray}

Performing the change of variables $z^{\prime}=z-i\hat{\tilde{\lambda}}\frac{\Delta S\sqrt{\sigma_{\xi}^{2}}}{\sqrt{q-\tilde{S}}}$,
where we set $\Delta S=S-\tilde{S}$, and two rotations between ($\tilde{z}$,
$x$) and ($z$, $x$) one can isolate the dependence over $x$ and
compute the $\int\mathcal{D}x$ integral analytically:

\begin{eqnarray*}
 &  & \int\mathcal{D}x\,H\left(\frac{K-\overline{\xi}\tilde{M}-\tilde{z}\sqrt{\sigma_{\xi}^{2}\tilde{q}}-z\left(\frac{\Delta S\sqrt{\sigma_{\xi}^{2}\tilde{q}}}{\sqrt{q\tilde{q}-\tilde{S}^{2}}}\right)-x\left(\frac{\Delta S\sqrt{\sigma_{\xi}^{2}\left(\tilde{q}-\tilde{S}\right)\tilde{S}}}{\sqrt{\left(q\tilde{q}-\tilde{S}^{2}\right)\left(q-\tilde{S}\right)}}\right)}{\sqrt{\sigma_{\xi}^{2}\left(\tilde{Q}-\tilde{q}-\frac{\left(\Delta S\right)^{2}}{q-\tilde{S}}\right)}}\right)=\\
 &  & =H\left(\frac{K-\overline{\xi}\tilde{M}-\tilde{z}\sqrt{\sigma_{\xi}^{2}\tilde{q}}-z\left(\frac{\Delta S\sqrt{\sigma_{\xi}^{2}\tilde{q}}}{\sqrt{q\tilde{q}-\tilde{S}^{2}}}\right)}{\sqrt{\sigma_{\xi}^{2}\left(\tilde{Q}-\tilde{q}-\frac{\left(\Delta S\right)^{2}}{q-\tilde{S}}+\frac{\Delta S^{2}\tilde{S}\left(\tilde{q}-\tilde{S}\right)}{\left(q\tilde{q}-\tilde{S}^{2}\right)\left(q-\tilde{S}\right)}\right)}}\right)
\end{eqnarray*}

now we can integrate over $\hat{\tilde{\lambda}}$ and $\tilde{\lambda}$
to obtain:

\begin{eqnarray}
 &  & \mathcal{G}_{E}=\\
 &  & =\left\langle \int\mathcal{D}z\int\mathcal{D}\tilde{z}\frac{H\left(\frac{K-s\overline{\xi}\tilde{M}-\tilde{z}\sqrt{\sigma_{\xi}^{2}\tilde{q}}-z\left(\frac{\Delta S\sqrt{\sigma_{\xi}^{2}\tilde{q}}}{\sqrt{q\tilde{q}-\tilde{S}^{2}}}\right)}{\sqrt{\sigma_{\xi}^{2}\left(\tilde{Q}-\tilde{q}-\frac{\left(\Delta S\right)^{2}}{q-\tilde{S}}+\frac{\Delta S^{2}\tilde{S}\left(\tilde{q}-\tilde{S}\right)}{\left(q\tilde{q}-\tilde{S}^{2}\right)\left(q-\tilde{S}\right)}\right)}}\right)\log\left(H\left(\frac{K-s\overline{\xi}M-z\sqrt{\sigma_{\xi}^{2}\left(\frac{q\tilde{q}-\tilde{S}^{2}}{\tilde{q}}\right)}-\tilde{z}\sqrt{\sigma_{\xi}^{2}\frac{\tilde{S}^{2}}{\tilde{q}}}}{\sqrt{\sigma_{\xi}^{2}\left(Q-q\right)}}\right)\right)}{H\left(\frac{K-s\overline{\xi}\tilde{M}-\tilde{z}\sqrt{\sigma_{\xi}^{2}\tilde{q}}}{\sqrt{\sigma_{\xi}^{2}\left(\tilde{Q}-\tilde{q}\right)}}\right)}\right\rangle _{s}\nonumber 
\end{eqnarray}
Plugging all the terms into the expression of the volume, we can now
write a saddle point approximation for the local entropy $\Phi\left(D\right)$:

\begin{eqnarray}
 & \mathcal{S}_{FP}\left(D\right)\,\,\,\,\text{\ensuremath{\approx}}\,\,\,\,\frac{1}{2}\hat{q}q-\hat{Q}Q-\hat{M}\overline{W}-\hat{S}S+\hat{\tilde{S}}\tilde{S}-\hat{D}\left(\frac{1}{2}Q+\frac{1}{2}\tilde{Q}-S-2D\right)+\\
 & +\int\mathcal{D}z\int\mathcal{D}\tilde{z}\frac{\sum_{\tilde{l}}\exp\left(\left(\hat{\tilde{Q}}-\frac{1}{2}\hat{\tilde{q}}\right)\tilde{l}^{2}+\left(\hat{\tilde{M}}+\tilde{z}\sqrt{\hat{q}}\right)\tilde{l}\right)\log\left(\sum_{l}\exp\left(\left(\hat{Q}-\frac{1}{2}\hat{q}\right)l^{2}+\left(\hat{M}+z\sqrt{\frac{\hat{q}\hat{\tilde{q}}-\hat{\tilde{S}}^{2}}{\hat{\tilde{q}}}}+\tilde{z}\frac{\hat{\tilde{S}}}{\sqrt{\hat{\tilde{q}}}}+\Delta\hat{S}\,\,\tilde{l}\right)l\right)\right)}{\sum_{\tilde{l}}\exp\left(\left(\hat{\tilde{Q}}-\frac{1}{2}\hat{\tilde{q}}\right)\tilde{l}^{2}+\left(\hat{\tilde{M}}+\tilde{z}\sqrt{\hat{q}}\right)\tilde{l}\right)}+\nonumber \\
 & +\alpha\left\langle \int\mathcal{D}z\int\mathcal{D}\tilde{z}\frac{H\left(\frac{K-s\overline{\xi}\tilde{M}-\tilde{z}\sqrt{\sigma_{\xi}^{2}\tilde{q}}-z\left(\frac{\Delta S\sqrt{\sigma_{\xi}^{2}\tilde{q}}}{\sqrt{q\tilde{q}-\tilde{S}^{2}}}\right)}{\sqrt{\sigma_{\xi}^{2}\left(\tilde{Q}-\tilde{q}-\frac{\left(\Delta S\right)^{2}}{q-\tilde{S}}+\frac{\Delta S^{2}\tilde{S}\left(\tilde{q}-\tilde{S}\right)}{\left(q\tilde{q}-\tilde{S}^{2}\right)\left(q-\tilde{S}\right)}\right)}}\right)\log\left(H\left(\frac{K-s\overline{\xi}M-z\sqrt{\sigma_{\xi}^{2}\left(\frac{q\tilde{q}-\tilde{S}^{2}}{\tilde{q}}\right)}-\tilde{z}\sqrt{\sigma_{\xi}^{2}\frac{\tilde{S}^{2}}{\tilde{q}}}}{\sqrt{\sigma_{\xi}^{2}\left(Q-q\right)}}\right)\right)}{H\left(\frac{K-s\overline{\xi}\tilde{M}-\tilde{z}\sqrt{\sigma_{\xi}^{2}\tilde{q}}}{\sqrt{\sigma_{\xi}^{2}\left(\tilde{Q}-\tilde{q}\right)}}\right)}\right\rangle _{s}\nonumber 
\end{eqnarray}

where all the order parameters must satisfy the saddle point equations,
found by requiring the stationarity condition $\delta\mathcal{S}_{FP}=0$:

\begin{eqnarray}
 & q=-2\frac{\partial}{\partial\hat{q}}\mathcal{G}_{S};\quad Q=\frac{\partial}{\partial\hat{Q}}\mathcal{G}_{S};\quad\overline{W}=\frac{\partial}{\partial\hat{M}}\mathcal{G}_{S};\quad\tilde{S}=\frac{\partial}{\partial\hat{\tilde{S}}}\mathcal{G}_{S};\quad S=\frac{Q}{2}+\frac{\tilde{Q}}{2}-2D;\quad0=\frac{\partial}{\partial\hat{S}}\mathcal{G}_{S}-S;\label{eq:FPsys}\\
 & \hat{q}=-2\alpha\frac{\partial}{\partial q}\mathcal{G}_{E};\quad\hat{Q}=-\frac{\hat{D}}{2}+\alpha\frac{\partial}{\partial Q}\mathcal{G}_{E};\quad\hat{D}=\hat{S}-\alpha\frac{\partial}{\partial S}\mathcal{G}_{E};\quad\hat{\tilde{S}}=-\alpha\frac{\partial}{\partial\tilde{S}}\mathcal{G}_{E};\quad\hat{M}=0;\quad0=\frac{\partial}{\partial M}\mathcal{G}_{E}.\nonumber 
\end{eqnarray}

Since the reference solution is sampled independently from the flat
Boltzmann distribution, the typical value for the order parameters
$\tilde{Q}$, $\tilde{q}$, and $\tilde{M}$ can be determined by
studying the simpler uncoupled replicated system: 

\begin{eqnarray*}
\tilde{\Omega}^{n} & = & \left\langle \int\prod_{i,c}d\mu\left(\tilde{W}_{i}^{c}\right)\prod_{c}\mathbb{X}_{\xi,\sigma}\left(\tilde{W}^{c},K\right)\right\rangle _{\xi,\sigma}
\end{eqnarray*}

In the end one can explicitly use the measure on the weights (eq.~\eqref{eq:weights_measure})
and get the saddle point equations:

\begin{eqnarray}
\tilde{q} & = & \int\mathcal{D}\tilde{z}\frac{\sum_{\tilde{l}}\left(\left(\tilde{l}^{2}-\frac{\tilde{l}\tilde{z}}{\sqrt{\hat{\tilde{q}}}}\right)\exp\left(\left(\hat{\tilde{Q}}-\frac{1}{2}\hat{\tilde{q}}\right)\tilde{l}^{2}+\left(\tilde{z}\sqrt{\hat{\tilde{q}}}\right)\tilde{l}\right)\right)}{\sum_{\tilde{l}}\exp\left(\left(\hat{\tilde{Q}}-\frac{1}{2}\hat{\tilde{q}}\right)\tilde{l}^{2}+\left(\tilde{z}\sqrt{\hat{\tilde{q}}}\right)\tilde{l}\right)}\\
\tilde{Q} & = & \int\mathcal{D}\tilde{z}\frac{\sum_{\tilde{l}}\left(\tilde{l}^{2}\exp\left(\left(\hat{\tilde{Q}}-\frac{1}{2}\hat{\tilde{q}}\right)\tilde{l}^{2}+\left(\tilde{z}\sqrt{\hat{\tilde{q}}}\right)\tilde{l}\right)\right)}{\sum_{\tilde{l}}\exp\left(\left(\hat{\tilde{Q}}-\frac{1}{2}\hat{\tilde{q}}\right)\tilde{l}^{2}+\left(\tilde{z}\sqrt{\hat{\tilde{q}}}\right)\tilde{l}\right)}\\
\overline{\tilde{W}} & = & \int\mathcal{D}\tilde{z}\frac{\sum_{\tilde{l}}\left(\tilde{l}\exp\left(\left(\hat{\tilde{Q}}-\frac{1}{2}\hat{\tilde{q}}\right)\tilde{l}^{2}+\left(\tilde{z}\sqrt{\hat{\tilde{q}}}\right)\tilde{l}\right)\right)}{\sum_{\tilde{l}}\exp\left(\left(\hat{\tilde{Q}}-\frac{1}{2}\hat{\tilde{q}}\right)\tilde{l}^{2}+\left(\tilde{z}\sqrt{\hat{\tilde{q}}}\right)\tilde{l}\right)}\\
0 & = & \left\langle \int\mathcal{D}\tilde{z}\,\mathcal{G}\left(\frac{K-s\overline{\xi}\tilde{M}-\tilde{z}\sqrt{\sigma_{\xi}^{2}\tilde{q}}}{\sqrt{\sigma_{\xi}^{2}\left(\tilde{Q}-\tilde{q}\right)}}\right)\right\rangle _{s}\label{eq:homo1}\\
\hat{\tilde{q}} & = & \alpha\left\langle \int\mathcal{D}\tilde{z}\,\mathcal{G}\left(\frac{K-s\overline{\xi}\tilde{M}-\tilde{z}\sqrt{\sigma_{\xi}^{2}\tilde{q}}}{\sqrt{\sigma_{\xi}^{2}\left(\tilde{Q}-\tilde{q}\right)}}\right)\left(-\frac{\tilde{z}}{\sqrt{\tilde{q}\left(\tilde{Q}-\tilde{q}\right)}}+\frac{K-s\overline{\xi}\tilde{M}-\tilde{z}\sqrt{\sigma_{\xi}^{2}\tilde{q}}}{\sqrt{\sigma_{\xi}^{2}}\left(\tilde{Q}-\tilde{q}\right)^{3/2}}\right)\right\rangle _{s}\\
\hat{\tilde{Q}} & = & \alpha\left\langle \int\mathcal{D}\tilde{z}\,\mathcal{G}\left(\frac{K-s\overline{\xi}\tilde{M}-\tilde{z}\sqrt{\sigma_{\xi}^{2}\tilde{q}}}{\sqrt{\sigma_{\xi}^{2}\left(\tilde{Q}-\tilde{q}\right)}}\right)\left(\frac{1}{2}\frac{K-s\overline{\xi}\tilde{M}-\tilde{z}\sqrt{\sigma_{\xi}^{2}\tilde{q}}}{\sqrt{\sigma_{\xi}^{2}}\left(\tilde{Q}-\tilde{q}\right)^{3/2}}\right)\right\rangle _{s}\\
\hat{\tilde{M}} & = & 0
\end{eqnarray}

where we defined $\mathcal{G}\left(x\right)=\frac{1}{H\left(x\right)}\frac{e^{-\frac{x^{2}}{2}}}{\sqrt{2\pi}}$.

This sub-system of 7 coupled equations can be easily solved iteratively
for each value of the control parameter $\alpha$, using Newton's
method for the implicit equation~\eqref{eq:homo1}. Then the saddle
point solutions can be substituted in the system~\eqref{eq:FPsys},
where there is the additional control parameter $D$. In order to
minimize the number of remaining implicit equations and help convergence
one can alternatively recast the equations and use $\hat{Q}$ as a
control parameter, since at the saddle point it is a bijective function
of the distance. Again, the saddle point solutions can be found by
iterating and using Newon's method.

\section{Reweighted measure, Constrained case\label{AppendixB:Modified-measure}}

Since we are looking for highly dense regions of solutions, we need
to consider a model where the statistical measure is reweighted in
order to increase the contribution of individual solutions surrounded
by a large number of other solutions. We study the large-deviation
free entropy density: 
\begin{equation}
\Phi_{RC}\left(D,y\right)=\frac{1}{N}\left\langle \log\left(\sum_{\left\{ \tilde{W}\right\} }\mathbb{X}_{\xi,\sigma}\left(\tilde{W},K\right)\,\mathcal{N}\left(\tilde{W},D\right)^{y}\right)\right\rangle _{\xi,\sigma}\equiv\frac{1}{N}\lim_{n\to0}\frac{\partial}{\partial n}\Omega_{RC}^{n}\left(D,y\right)\label{eq:free_energy_finite_y}
\end{equation}
where the inverse temperature $y$ can be used to focus on these regions,
$\mathbb{X}_{\xi,\sigma}\left(W,K\right)$ is defined as in \eqref{eq:indicator_withK}
and $\mathcal{N}\left(\tilde{W},D\right)=\sum_{\left\{ W\right\} }\mathbb{X}_{\xi,\sigma}\left(W,K\right)\delta\left(d\left(W,\tilde{W}\right)-D\right)$
is the number of solutions at distance $D$ from the reference configuration.
In the case $y=0$ we recover the standard case in which the measure
is flat (denoted as $\Phi$ in the main text).

Like in the previous calculation we can evaluate the quenched average
over the set of patterns $\left\{ \xi^{\mu},\sigma^{\mu}\right\} _{\mu=1,...,\alpha N}$
by exploiting the replica trick: in this picture the $y$ temperature
can be formally interpreted as the number of auxiliary configurations
$W$ assigned to each reference configuration $\tilde{W}$.

In the following the indices $c,d\in\left\{ 1,\dots,n\right\} $ will
denote the number of replicas of the reference configurations, while
the $yn$ auxiliary replicas will be denoted also by the indices $a,b\in\left\{ 1,\dots,y\right\} $.
Only the $n\to0$ limit must be taken while $y$ will remain as a
parameter of the problem.

We thus need to evaluate the replicated volume (we use the measure
of eq.~\eqref{eq:weights_measure} on the weights):

\begin{eqnarray}
 &  & \Omega_{RC}^{n}\left(D,y\right)=\\
 &  & =\left<\int\prod_{i,c}d\mu\left(\tilde{W}_{i}^{c}\right)\int\prod_{i,ca}d\mu\left(W_{i}^{ca}\right)\prod_{c}\mathbb{X}_{\xi,\sigma}\left(\tilde{W}^{c},K\right)\prod_{ca}\mathbb{X}_{\xi,\sigma}\left(W^{ca},K\right)\times\right.\nonumber \\
 &  & \qquad\left.\times\prod_{ca}\delta\left(\frac{1}{2}\sum_{i}\left(W_{i}^{ca}-\tilde{W_{i}^{c}}\right)^{2}-2DN\right)\right>_{\xi,\sigma}\nonumber \\
 &  & =\int\prod_{i,c}d\mu\left(\tilde{W}_{i}^{c}\right)\int\prod_{i,ca}d\mu\left(W_{i}^{ca}\right)\prod_{ca}\delta\left(\frac{1}{2}\sum_{i}\left(W_{i}^{ca}\right)^{2}+\frac{1}{2}\sum_{i}\left(\tilde{W}_{i}^{c}\right)^{2}-\sum_{i}W_{i}^{ca}\tilde{W_{i}^{c}}-2DN\right)\times\nonumber \\
 &  & \qquad\times\int\prod_{\mu,c}\frac{d\tilde{\lambda}_{\mu}^{c}d\hat{\tilde{\lambda}}_{\mu}^{c}}{2\pi}\prod_{\mu,c}e^{i\tilde{\lambda}_{\mu}^{c}\hat{\tilde{\lambda}}_{\mu}^{c}}\int\prod_{\mu,ca}\frac{d\lambda_{\mu}^{ca}d\hat{\lambda}_{\mu}^{ca}}{2\pi}\left\langle \prod_{\mu,c}\Theta\left(s^{\mu}\tilde{\lambda}_{\mu}^{c}-K\right)\prod_{\mu,ca}\Theta\left(s^{\mu}\lambda_{\mu}^{ca}-K\right)\right\rangle _{s^{\mu}}\times\nonumber \\
 &  & \qquad\times\prod_{\mu,ca}e^{i\lambda_{\mu}^{ca}\hat{\lambda}_{\mu}^{ca}}\prod_{\mu,i}\left(e^{i\theta\sqrt{N}\left(\sum_{c}\hat{\tilde{\lambda}}_{\mu}^{c}+\sum_{ac}\hat{\lambda}_{\mu}^{ca}\right)}\left\langle \exp\left(-\frac{i}{\sqrt{N}}\left(\sum_{c}\hat{\tilde{\lambda}}_{\mu}^{c}\tilde{W}_{i}^{c}+\sum_{ca}\hat{\lambda}_{\mu}^{ca}W_{i}^{ca}\right)\xi_{i}^{\mu}\right)\right\rangle _{\xi_{i}^{\mu}}\right)\nonumber 
\end{eqnarray}
where we used the auxiliary output variables $s^{\mu}$ (see eq.~\eqref{eq:output_average})
instead of the $\sigma^{\mu}$, and we substituted the arguments of
the indicator functions in order to perform the average over the inputs
as in \eqref{eq:disorder_av}:

\begin{eqnarray}
 &  & \prod_{\mu,i}\left\langle \exp\left(-\frac{i}{\sqrt{N}}\left(\sum_{c}\hat{\tilde{\lambda}}_{\mu}^{c}\tilde{W}_{i}^{c}+\sum_{ca}\hat{\lambda}_{\mu}^{ca}W_{i}^{ca}\right)\xi_{i}^{\mu}\right)\right\rangle _{\xi}=\\
 &  & =\prod_{\mu}\exp\left(-i\overline{\xi}\sqrt{N}\left(\sum_{c}\hat{\tilde{\lambda}}_{\mu}^{c}\sum_{i}\frac{\tilde{W}_{i}^{c}}{N}+\sum_{ca}\hat{\lambda}_{\mu}^{ca}\sum_{i}\frac{W_{i}^{ca}}{N}\right)+\right.\nonumber \\
 &  & \left.\qquad\qquad-\frac{1}{2}\sigma_{\xi}^{2}\left(\sum_{ca,db}\hat{\lambda}_{\mu}^{ca}\hat{\lambda}_{\mu}^{db}\sum_{i}\frac{W_{i}^{ca}W_{i}^{db}}{N}+\sum_{c,d}\hat{\tilde{\lambda}}_{\mu}^{c}\hat{\tilde{\lambda}}_{\mu}^{d}\sum_{i}\frac{\tilde{W}_{i}^{c}\tilde{W}_{i}^{d}}{N}+2\sum_{ca,d}\hat{\lambda}_{\mu}^{ca}\hat{\tilde{\lambda}}_{\mu}^{d}\sum_{i}\frac{W_{i}^{ca}\tilde{W}_{i}^{d}}{N}\right)\right)\nonumber 
\end{eqnarray}

Using the same notation as in the previous section we substitute all
the obtained overlaps, defining the following order parameters, fixed
by introducing the related Dirac $\delta$ distributions:
\begin{itemize}
\item $\sum_{i}\frac{\left(\tilde{W}_{i}\right)^{2}}{N}=\tilde{Q}$ , $\sum_{i}\frac{\left(W_{i}^{ca}\right)^{2}}{N}=Q^{ca}$ 
\item $\sum_{i}\frac{\tilde{W}_{i}^{c}}{N}=\overline{\tilde{W}}+\frac{\tilde{M}^{c}}{\sqrt{N}}$,
$\sum_{i}\frac{W_{i}^{ca}}{N}=\overline{W}+\frac{M^{ca}}{\sqrt{N}}$
\item $\sum_{i}\frac{\tilde{W}_{i}^{c}\tilde{W}_{i}^{d}}{N}=\tilde{q}^{cd}$
, $\sum_{i}\frac{W_{i}^{ca}W_{i}^{db}}{N}=q^{ca,db}$ ,$\sum_{i}\frac{W_{i}^{ca}\tilde{W}_{i}^{d}}{N}=S^{ca,d}$ 
\end{itemize}
After the substitutions in the expression of the volume one gets:

\begin{eqnarray}
\Omega_{RC}^{n}\left(D,y\right) & = & \int\prod_{_{c>d,ab}^{c,a>b}}\frac{dq^{ca,db}d\hat{q}^{ca,db}}{\left(2\pi/N\right)}\int\prod_{c>d}\frac{d\tilde{q}^{cd}d\hat{\tilde{q}}^{cd}}{\left(2\pi/N\right)}\int\prod_{ca}\frac{dQ^{ca}d\hat{Q}^{ca}}{\left(2\pi/N\right)}\int\prod_{c}\frac{d\tilde{Q}^{c}d\hat{\tilde{Q}}^{c}}{\left(2\pi/N\right)}\label{eq:before_ansatz_RS}\\
 &  & \int\prod_{ca}\frac{dM^{ca}d\hat{M}^{ca}}{\left(2\pi/\sqrt{N}\right)}\int\prod_{c}\frac{d\tilde{M}^{c}d\hat{\tilde{M}}^{c}}{\left(2\pi/\sqrt{N}\right)}\int\prod_{ca,d}\frac{dS^{ca,d}d\hat{S}^{ca,d}}{\left(2\pi/N\right)}\int\prod_{ca}\frac{d\hat{D}^{ca}}{2\pi}\,G_{1}\,G_{S}^{N}\,G_{E}^{\alpha N}\nonumber 
\end{eqnarray}
where, as in the previous section, we could factorize over the indices
$\mu$ and $i$ (thus removing all those indices) and we defined: 

\begin{eqnarray}
 & G_{1}= & \exp\left(-N\left(\sum_{c}\sum_{a>b}\hat{q}^{ca,cb}q^{ca,cb}+\sum_{c>d}\sum_{ab}\hat{q}^{ca,db}q^{ca,db}+\sum_{c>d}\hat{\tilde{q}}^{cd}\tilde{q}^{cd}+\right.\right.\\
 &  & +\sum_{ca}\hat{Q}^{ca}Q^{ca}+\sum_{c}\hat{\tilde{Q}}^{c}\tilde{Q}^{c}+\sum_{c}\hat{\tilde{M}}^{c}\left(\frac{\tilde{M}^{c}}{\sqrt{N}}+\overline{W}\right)+\sum_{ca}\hat{M}^{ca}\left(\frac{M^{ca}}{\sqrt{N}}+\overline{W}\right)+\sum_{ca,d}\hat{S}^{ca,d}S^{ca,d}+\nonumber \\
 &  & \left.\left.+\sum_{ca}\hat{D}^{ca}\left(\frac{1}{2}Q^{ca}+\frac{1}{2}\tilde{Q}^{c}-S^{ca,c}-2D\right)\right)\right)\nonumber \\
 & G_{S}= & \int\prod_{c}d\mu\left(\tilde{W}^{c}\right)\int\prod_{ca}d\mu\left(W^{ca}\right)\exp\left(\sum_{c}\sum_{a>b}\hat{q}^{ca,cb}W^{ca}W^{cb}+\sum_{c>d}\sum_{ab}\hat{q}^{ca,db}W^{ca}W^{db}+\right.\\
 &  & +\sum_{c>d}\hat{\tilde{q}}^{cd}\tilde{W}_{i}^{c}\tilde{W}_{i}^{d}+\sum_{ca}\hat{Q}^{ca}\left(W^{ca}\right)^{2}+\sum_{c}\hat{\tilde{Q}}^{c}\left(\tilde{W}^{c}\right)^{2}+\sum_{ca}\hat{M}^{ca}W^{ca}+\sum_{c}\hat{\tilde{M}}^{c}\tilde{W}^{c}+\nonumber \\
 &  & \left.+\sum_{dca}\hat{S}^{d,ca}W^{ca}\tilde{W}^{d}\right)\nonumber \\
 & G_{E}= & \int\prod_{c}\frac{d\tilde{\lambda}^{c}d\hat{\tilde{\lambda}}^{c}}{2\pi}\int\prod_{ca}\frac{d\lambda^{ca}d\hat{\lambda}^{ca}}{2\pi}\left\langle \prod_{c}\Theta\left(s\tilde{\lambda}^{c}-K\right)\prod_{\mu,ca}\Theta\left(s\lambda^{ca}-K\right)\right\rangle _{s}\times\\
 &  & \times\exp\left(i\left(\sum_{c}\tilde{\lambda}^{c}\hat{\tilde{\lambda}}^{c}+\sum_{ca}\lambda^{ca}\hat{\lambda}^{ca}-\overline{\xi}\left(\sum_{ca}\hat{\lambda}^{ca}M^{ca}-\sum_{c}\hat{\tilde{\lambda}}^{c}\tilde{M}^{c}\right)\right)\right)\times\nonumber \\
 &  & \times\exp\left(\sigma_{\xi}^{2}\left(-\frac{1}{2}\sum_{c}\left(\hat{\tilde{\lambda}}^{c}\right)^{2}\tilde{Q}^{c}-\frac{1}{2}\sum_{ca}\left(\hat{\lambda}^{ca}\right)^{2}Q^{ca}-\sum_{c>d}\hat{\tilde{\lambda}}^{c}\hat{\tilde{\lambda}}^{d}\tilde{q}^{cd}+\right.\right.\nonumber \\
 &  & \left.\left.\qquad-\sum_{c}\sum_{a>b}\hat{\lambda}^{ca}\hat{\lambda}^{cb}q^{ca,cb}-\sum_{c>d}\sum_{ab}\hat{\lambda}^{ca}\hat{\lambda}^{db}q^{ca,db}-\sum_{ca,d}\hat{\lambda}^{ca}\hat{\tilde{\lambda}}^{d}S^{d,ca}\right)\right)\nonumber 
\end{eqnarray}

\subsection{Replica Symmetric Ansatz	}

As in the Franz-Parisi analysis we now need to make a simplification,
putting forward an Ansatz on the structure of the parameters describing
the replicated system. We start from a replica symmetric Ansatz; notice
however that the reweighting term already introduced a natural grouping
of the students $W$ in sets of $y$ elements, each surrounding a
certain reference solution $\tilde{W}$, thus leading to a situation
formally similar to a 1RSB description.

Therefore we have to make a distinction between the typical overlap
$q_{1}$, between replicas found around the same $\tilde{W}$, and
the overlap $q_{0}$, between replicas referred to different ones:
\begin{itemize}
\item $q^{ca,cb}=q_{1}$ for $\left(a\neq b\right)$, $q^{ca,db}=q_{0}$
for $\left(c\neq d\right)$ 
\item $S^{ca,c}=S$, $S^{ca,d}=\tilde{S}$ for $\left(c\neq d\right)$ 
\item $\tilde{Q}^{c}=\tilde{Q}$ , $Q^{ca}=Q$ , $\tilde{M}^{c}=M$ , $M^{ca}=M$
, $\tilde{q}^{cd}=\tilde{q}$ , $\hat{D}^{ca}=\hat{D}$ 
\end{itemize}
With these assumptions we can proceed in the computation of the replicated
volume. 

First, neglecting the $O\left(n^{2}\right)$ terms, we find for $G_{1}$:

\begin{eqnarray}
G_{1} & = & \exp\left(-N\left(n\,\frac{y\left(y-1\right)}{2}\,\hat{q}_{1}q_{1}+\frac{n\left(n-1\right)}{2}\,y^{2}\,\hat{q}_{0}q_{0}+\frac{n\left(n-1\right)}{2}\hat{\tilde{q}}\tilde{q}+ny\hat{Q}Q+n\hat{\tilde{Q}}\tilde{Q}+\right.\right.\\
 &  & \left.\left.+ny\hat{M}\overline{W}+n\hat{\tilde{M}}\overline{\tilde{W}}+ny\hat{S}S+\frac{n\left(n-1\right)}{2}y\hat{\tilde{S}}\tilde{S}+ny\hat{D}\left(\frac{1}{2}Q+\frac{1}{2}\tilde{Q}-S-2D\right)\right)\right)\nonumber \\
 & = & \exp\left(-Nny\left.\frac{\left(y-1\right)}{2}q_{1}q_{1}-\frac{y}{2}\hat{q}_{0}q_{0}-\frac{1}{2}\frac{\hat{\tilde{q}}\tilde{q}}{y}+\hat{Q}Q+\frac{\hat{\tilde{Q}}\tilde{Q}}{y}+\hat{M}\overline{W}+\right.\right.\nonumber \\
 &  & \left.+\frac{\hat{\tilde{M}}\overline{\tilde{W}}}{y}+\hat{S}S-\hat{\tilde{S}}\tilde{S}+\hat{D}\left(\frac{1}{2}Q+\frac{1}{2}\tilde{Q}-S-2D\right)\right)\nonumber 
\end{eqnarray}

In the computation of the entropic term we follow closely the steps
explained in the previous section.

We recast $\hat{\tilde{S}}\sum_{ca}W^{ca}\sum_{c}\tilde{W}^{c}=\frac{1}{2}\hat{\tilde{S}}\left(\sum_{ca}W^{ca}+\sum_{c}\tilde{W}^{c}\right)^{2}-\frac{1}{2}\hat{\tilde{S}}\left(\sum_{ca}W^{ca}\right)^{2}-\frac{1}{2}\hat{\tilde{S}}\left(\sum_{c}\tilde{W}^{c}\right)^{2}$,
we then introduce the variables $x$, $z_{0}$, $\tilde{z}$ to perform
three Hubbard-Stratonovich transformations, thus getting rid of the
squared sums involving the replica index $c$ and factorizing over
it: 

\begin{eqnarray}
G_{S} & = & \int\mathcal{D}x\int\mathcal{D}z_{0}\int\mathcal{D}\tilde{z}\left\{ \int d\mu\left(\tilde{W}\right)\int\prod_{a}d\mu\left(W^{a}\right)\exp\left(\left(\hat{Q}-\frac{1}{2}\hat{q}_{1}\right)\sum_{a}(W^{a})^{2}+\right.\right.\\
 &  & +\frac{1}{2}(\hat{q}_{1}-\hat{q}_{0})\left(\sum_{a}W^{a}\right)^{2}+z_{0}\sqrt{\hat{q}_{0}-\hat{\tilde{S}}}\sum_{a}W^{a}+\hat{M}\sum_{a}W^{a}+\left(\hat{\tilde{Q}}-\frac{1}{2}\tilde{q}\right)\tilde{W}^{2}+\tilde{z}\sqrt{\hat{\tilde{q}}-\hat{\tilde{S}}}\tilde{W}+\nonumber \\
 &  & \left.\left.+\hat{\tilde{M}}\tilde{W}+\left(\hat{S}-\hat{\tilde{S}}\right)\sum_{a}W^{a}\tilde{W}+x\sqrt{\hat{\tilde{S}}}\sum_{a}W^{a}+x\sqrt{\hat{\tilde{S}}}\tilde{W}\right)\right\} ^{n}\nonumber 
\end{eqnarray}

Now we perform the last Hubbard-Stratonovich transformation and factorize
over the index $a$ as well, obtaining:

\begin{eqnarray}
G_{S} & = & \int\mathcal{D}x\int\mathcal{D}z_{0}\int\mathcal{D}\tilde{z}\left\{ \int d\mu\left(\tilde{W}\right)\exp\left(\left(\hat{\tilde{Q}}-\frac{1}{2}\tilde{q}\right)\tilde{W}^{2}+\left(\tilde{z}\sqrt{\hat{\tilde{q}}-\hat{\tilde{S}}}+\hat{\tilde{M}}+x\sqrt{\hat{\tilde{S}}}\right)\tilde{W}\right)\times\right.\\
 &  & \times\int\mathcal{D}z_{1}\left\{ \int d\mu\left(W\right)\exp\left(\left(\hat{Q}-\frac{1}{2}\hat{q}_{1}\right)W^{2}+\right.\right.\nonumber \\
 &  & \qquad\left.\left.\left.+\left(z_{0}\sqrt{\hat{q}_{0}-\hat{\tilde{S}}}+z_{1}\sqrt{\hat{q}_{1}-\hat{q}_{0}}+x\sqrt{\hat{\tilde{S}}}+\hat{M}+\left(\hat{S}-\hat{\tilde{S}}\right)\tilde{W}\right)W\right)\right\} ^{y}\right\} ^{n}
\end{eqnarray}

and in the limit $n\to0$ (and explicitly using the measure of eq.~\eqref{eq:weights_measure}):

\begin{eqnarray}
 &  & \mathcal{G}_{S}=\frac{1}{n}\log G_{S}=\\
 &  & =\int\mathcal{D}x\int\mathcal{D}z_{0}\int\mathcal{D}\tilde{z}\log\left\{ \sum_{\tilde{l}}\exp\left(\left(\hat{\tilde{Q}}-\frac{1}{2}\tilde{q}\right)\tilde{l}^{2}+\left(\tilde{z}\sqrt{\hat{\tilde{q}}-\hat{\tilde{S}}}+\hat{\tilde{M}}+x\sqrt{\hat{\tilde{S}}}\right)\tilde{l}\right)\times\right.\nonumber \\
 &  & \quad\left.\times\int\mathcal{D}z_{1}\left\{ \sum_{l}\exp\left(\left(\hat{Q}-\frac{1}{2}\hat{q}_{1}\right)l^{2}+\left(z_{0}\sqrt{\hat{q}_{0}-\hat{\tilde{S}}}+z_{1}\sqrt{\hat{q}_{1}-\hat{q}_{0}}+x\sqrt{\hat{\tilde{S}}}+\hat{M}+\left(\hat{S}-\hat{\tilde{S}}\right)\tilde{l}\right)l\right)\right\} ^{y}\right\} \nonumber 
\end{eqnarray}

Then we perform two changes of variables in order to evaluate analytically
the $\int\mathcal{D}x$ integral and obtain:

\begin{eqnarray}
\mathcal{G}_{S} & = & \int\mathcal{D}\tilde{z}\int\mathcal{D}z_{0}\log\left\{ \sum_{\tilde{l}}\exp\left(\left(\hat{\tilde{Q}}-\frac{1}{2}\tilde{q}\right)\tilde{l}^{2}+\left(\tilde{z}\sqrt{\hat{\tilde{q}}-\frac{\hat{\tilde{S}}^{2}}{\hat{q}_{0}}}+z_{0}\frac{\hat{\tilde{S}}}{\sqrt{\hat{q}_{0}}}+\hat{\tilde{M}}\right)\tilde{l}\right)\times\right.\\
 &  & \quad\left.\times\int\mathcal{D}z_{1}\left\{ \sum_{l}\exp\left(\left(\hat{Q}-\frac{1}{2}\hat{q}_{1}\right)l^{2}+\left(z_{0}\sqrt{\hat{q}_{0}}+z_{1}\sqrt{\hat{q}_{1}-\hat{q}_{0}}+\hat{M}+\left(\hat{S}-\hat{\tilde{S}}\right)\tilde{l}\right)l\right)\right\} ^{y}\right\} \nonumber 
\end{eqnarray}

In a similar way, we reorganize the summations in the energetic term,
and after the substitution

\begin{eqnarray*}
\sigma_{\xi}^{2}\tilde{S}\sum_{ca}\hat{\lambda}^{ca}\sum_{c}\hat{\tilde{\lambda}}^{c} & = & \frac{1}{2}\sigma_{\xi}^{2}\tilde{S}\left(\left(\sum_{ca}\hat{\lambda}^{ca}+\sum_{c}\hat{\tilde{\lambda}}^{c}\right)^{2}-\left(\sum_{c}\hat{\tilde{\lambda}}^{c}\right)^{2}-\left(\sum_{ca}\hat{\lambda}^{ca}\right)^{2}\right)
\end{eqnarray*}
we have:

\begin{eqnarray}
G_{E} & = & \int\prod_{c}\frac{d\tilde{\lambda}^{c}d\hat{\tilde{\lambda}}^{c}}{2\pi}\int\prod_{ca}\frac{d\lambda^{ca}d\hat{\lambda}^{ca}}{2\pi}\left\langle \prod_{c}\Theta\left(s\tilde{\lambda}^{c}-K\right)\prod_{ca}\Theta\left(s\lambda^{ca}-K\right)\right\rangle _{s}\times\\
 &  & \times\exp\left(i\left(\sum_{c}\tilde{\lambda}^{c}\hat{\tilde{\lambda}}^{c}+\sum_{ca}\lambda^{ca}\hat{\lambda}^{ca}-i\overline{\xi}\tilde{M}\sum_{c}\hat{\tilde{\lambda}}^{c}-i\overline{\xi}M\sum_{ca}\hat{\lambda}^{ca}\right)-\frac{\sigma_{\xi}^{2}}{2}\left(q_{0}-\tilde{S}\right)\left(\sum_{ca}\hat{\lambda}^{ca}\right)^{2}\right)\times\nonumber \\
 &  & \times\exp\left\{ -\frac{\sigma_{\xi}^{2}}{2}\left(\left(q_{1}-q_{0}\right)\sum_{c}\left(\sum_{a}\hat{\lambda}^{ca}\right)^{2}+\left(Q-q_{1}\right)\sum_{ca}\left(\hat{\lambda}^{ca}\right)^{2}+\left(\tilde{q}-\tilde{S}\right)\left(\sum_{c}\hat{\tilde{\lambda}}^{c}\right)^{2}+\right.\right.\nonumber \\
 &  & \qquad\left.\left.+\left(\tilde{Q}-\tilde{q}\right)\sum_{c}\left(\hat{\tilde{\lambda}}^{c}\right)^{2}-2\left(S-\tilde{S}\right)\sum_{ca}\hat{\lambda}^{ca}\hat{\tilde{\lambda}}^{c}-\tilde{S}\left(\sum_{ca}\hat{\lambda}^{ca}+\sum_{c}\hat{\tilde{\lambda}}^{c}\right)^{2}\right)\right\} \nonumber 
\end{eqnarray}

Again we perform three Hubbard-Stratonovich transformations, introducing
$x$, $z_{0}$, and $\tilde{z}$, and factorize over the index $c$.
Then we evaluate the Gaussian integral in the variable $\hat{\tilde{\lambda}}$,
getting: 

\begin{eqnarray}
G_{E} & = & \int\mathcal{D}x\int\mathcal{D}z_{0}\int\mathcal{D}\tilde{z}\left<\left\{ \int\frac{d\tilde{\lambda}}{\sqrt{2\pi}}\Theta\left(s\tilde{\lambda}-K\right)\times\right.\right.\\
 &  & \quad\times\exp\left(-\frac{1}{2}\frac{\left(\tilde{\lambda}-\overline{\xi}\tilde{M}-\tilde{z}\sqrt{\sigma_{\xi}^{2}\left(\tilde{q}-\tilde{S}\right)}-x\sqrt{\sigma_{\xi}^{2}\tilde{S}}+i\sigma_{\xi}^{2}\left(S-\tilde{S}\right)\hat{\lambda}\right)^{2}}{\sigma_{\xi}^{2}(\tilde{Q}-\tilde{q})}\right)\times\nonumber \\
 &  & \quad\times\int\prod_{a}\frac{d\lambda^{a}d\hat{\lambda}^{a}}{2\pi}\prod_{a}\Theta\left(s\lambda^{a}-K\right)\exp\left(i\left(\sum_{a}\lambda^{a}\hat{\lambda}^{a}-\left(\overline{\xi}M+z_{0}\sqrt{\sigma_{\xi}^{2}\left(q_{0}-\tilde{S}\right)}+x\sqrt{\sigma_{\xi}^{2}\tilde{S}}\right)\sum_{a}\hat{\lambda}^{a}\right)\right)\nonumber \\
 &  & \qquad\left.\left.\times\exp\left(-\frac{\sigma_{\xi}^{2}}{2}\left(\left(q_{1}-q_{0}\right)\left(\sum_{a}\hat{\lambda}^{a}\right)^{2}+(Q-q_{1})\sum_{a}\left(\hat{\lambda}^{a}\right)^{2}\right)\right)\right\} ^{n}\right>_{s}\nonumber 
\end{eqnarray}

We can define:

\begin{eqnarray}
\tilde{A}\left(\tilde{\lambda},x,\tilde{z}\right) & = & \tilde{\lambda}-\overline{\xi}\tilde{M}-\tilde{z}\sqrt{\sigma_{\xi}^{2}\left(\tilde{q}-\tilde{S}\right)}-x\sqrt{\sigma_{\xi}^{2}\tilde{S}}
\end{eqnarray}

and after a fourth Hubbard-Stratonovich transformation, with the variable
$z_{1}$, we factorize also over $a$:

\begin{eqnarray}
G_{E} & = & \int\mathcal{D}x\int\mathcal{D}z_{0}\int\mathcal{D}\tilde{z}\left<\left\{ \int\frac{d\tilde{\lambda}}{\sqrt{2\pi}}\Theta\left(s\tilde{\lambda}-K\right)\exp\left(-\frac{1}{2}\frac{\tilde{A}\left(\tilde{\lambda},x,\tilde{z}\right)^{2}}{\sigma_{\xi}^{2}\left(\tilde{Q}-\tilde{q}\right)}\right)\times\right.\right.\\
 &  & \quad\times\int\mathcal{D}z_{1}\left\{ \int\frac{d\lambda d\hat{\lambda}}{2\pi}\Theta\left(s\lambda-K\right)\exp\left(-\frac{\sigma_{\xi}^{2}}{2}(Q-q_{1})\hat{\lambda}^{2}\right)\times\right.\nonumber \\
 &  & \qquad\times\exp\left(i\hat{\lambda}\left(\lambda-\overline{\xi}M-z_{0}\sqrt{\sigma_{\xi}^{2}\left(q_{0}-\tilde{S}\right)}-z_{1}\sqrt{\sigma_{\xi}^{2}\left(q_{1}-q_{0}-\frac{\left(S-\tilde{S}\right)^{2}}{\tilde{Q}-\tilde{q}}\right)}+\right.\right.\nonumber \\
 &  & \qquad\qquad\left.\left.\left.\left.\left.-x\sqrt{\sigma_{\xi}^{2}\tilde{S}}-\frac{S-\tilde{S}}{\tilde{Q}-\tilde{q}}\tilde{A}(\tilde{\lambda},x,\tilde{z})\right)\right)\right\} ^{y}\right\} ^{n}\right>_{s}\nonumber 
\end{eqnarray}

Now we evaluate also the $\hat{\lambda}$ Gaussian integral, obtaining:

\begin{eqnarray*}
G_{E} & = & \int\mathcal{D}x\int\mathcal{D}z_{0}\int\mathcal{D}\tilde{z}\left\langle \left\{ \int d\tilde{\lambda}\frac{\Theta\left(s\tilde{\lambda}-K\right)}{\sqrt{2\pi\left(\sigma_{\xi}^{2}\left(\tilde{Q}-\tilde{q}\right)\right)}}\exp\left(-\frac{1}{2}\frac{\tilde{A}\left(\tilde{\lambda},x,\tilde{z}\right)^{2}}{\sigma_{\xi}^{2}\left(\tilde{Q}-\tilde{q}\right)}\right)\times\right.\right.\\
 &  & \quad\left.\left.\times\int\mathcal{D}z_{1}\left\{ \int\frac{d\lambda}{\sqrt{2\pi\left(\sigma_{\xi}^{2}\left(Q-q_{1}\right)\right)}}\Theta\left(s\lambda-K\right)\exp\left(-\frac{1}{2}\frac{A\left(\lambda,x,z_{0},z_{1},\tilde{z}\right)^{2}}{\sigma_{\xi}^{2}\left(Q-q_{1}\right)}\right)\right\} ^{y}\right\} ^{n}\right\rangle _{s}
\end{eqnarray*}

where we defined, after a rotation between $z_{0}$ and $x$:

\begin{eqnarray}
A\left(\lambda,x,z_{0},z_{1},\tilde{z}\right) & = & \lambda-\overline{\xi}M-z_{0}\sqrt{\sigma_{\xi}^{2}q_{0}}-z_{1}\sqrt{\sigma_{\xi}^{2}\left(q_{1}-q_{0}-\frac{\left(S-\tilde{S}\right)^{2}}{\tilde{Q}-\tilde{q}}\right)}-\frac{S-\tilde{S}}{\tilde{Q}-\tilde{q}}\tilde{A}^{\prime}\left(\tilde{\lambda},x,\tilde{z},z_{0}\right)\\
\tilde{A}^{\prime}\left(\tilde{\lambda},x,\tilde{z},z_{0}\right) & = & \tilde{\lambda}-\overline{\xi}\tilde{M}-\tilde{z}\sqrt{\sigma_{\xi}^{2}\left(\tilde{q}-\tilde{S}\right)}-z_{0}\sqrt{\sigma_{\xi}^{2}\frac{\tilde{S}^{2}}{q_{0}}}+x\sqrt{\sigma_{\xi}^{2}\frac{\tilde{S}\left(q_{0}-\tilde{S}\right)}{q_{0}}}
\end{eqnarray}

After a change in the sign of $x$ and another rotation between $\tilde{z}$
and $x$, we can simplify the integral in $x$ and perform a shift
in the variable $\tilde{\lambda}$, so to get:

\begin{eqnarray}
\tilde{\lambda}^{\prime} & = & \frac{\tilde{\lambda}-\overline{\xi}\tilde{M}-\tilde{z}\sqrt{\sigma_{\xi}^{2}\left(\tilde{q}-\frac{\tilde{S}^{2}}{q_{0}}\right)}-z_{0}\sqrt{\sigma_{\xi}^{2}\frac{\tilde{S}^{2}}{q_{0}}}}{\sqrt{\sigma_{\xi}^{2}\left(\tilde{Q}-\tilde{q}\right)}}\\
A\left(\lambda,z_{0},z_{1},\tilde{z}\right) & = & \lambda-\overline{\xi}M-z_{0}\sqrt{\sigma_{\xi}^{2}q_{0}}-z_{1}\sqrt{\sigma_{\xi}^{2}\left(q_{1}-q_{0}-\frac{\left(S-\tilde{S}\right)^{2}}{\tilde{Q}-\tilde{q}}\right)}-\sqrt{\sigma_{\xi}^{2}}\frac{S-\tilde{S}}{\sqrt{\tilde{Q}-\tilde{q}}}\tilde{\lambda}
\end{eqnarray}

Now we take the logarithm of the energetic term in the $n\to0$ limit,
and after rotating $z_{1}$ and $\tilde{\lambda}$ we can introduce
the integral functions defined in eq.~\eqref{eq:h_function} to finally
find:

\begin{eqnarray}
 &  & \mathcal{G}_{E}=\frac{1}{n}\log G_{E}=\\
 &  & =\int\mathcal{D}z_{0}\int\mathcal{D}\tilde{z}\,\left\langle \log\left(\int\mathcal{D}z_{1}H\left(\frac{K-s\overline{\xi}M-z_{0}\sqrt{\sigma_{\xi}^{2}q_{0}}-z_{1}\sqrt{\sigma_{\xi}^{2}\left(q_{1}-q_{0}\right)}}{\sqrt{\sigma_{\xi}^{2}\left(Q-q_{1}\right)}}\right)^{y}H\left(\tilde{C}\left(s,z_{0},z_{1},\tilde{z}\right)\right)\right)\right\rangle _{s}\nonumber 
\end{eqnarray}

with the definition:

\begin{eqnarray}
\tilde{C}\left(s,z_{0},z_{1},\tilde{z}\right) & = & \frac{K-s\overline{\xi}\tilde{M}-\tilde{z}\sqrt{\sigma_{\xi}^{2}\left(\tilde{q}-\frac{\tilde{S}^{2}}{q_{0}}\right)}-z_{0}\sqrt{\sigma_{\xi}^{2}\frac{\tilde{S}^{2}}{q_{0}}}-z_{1}^{'}\sqrt{\sigma_{\xi}^{2}}\frac{\left(S-\tilde{S}\right)}{\sqrt{\left(q_{1}-q_{0}\right)}}}{\sqrt{\sigma_{\xi}^{2}\left(\left(\tilde{Q}-\tilde{q}\right)-\frac{\left(S-\tilde{S}\right)^{2}}{\left(q_{1}-q_{0}\right)}\right)}}
\end{eqnarray}

\subsubsection{Final RS expression}

Putting the pieces together and using the saddle point method we finally
obtain a leading order estimate of the free energy density function
in the large $N$ limit:

\selectlanguage{british}%
\begin{eqnarray}
\Phi_{RC}\left(D,y\right) & \approx & -\left(\left(\frac{y}{2}\hat{q}_{1}q_{1}-\frac{y^{2}}{2}\left(\hat{q}_{1}q_{1}-\hat{q}_{0}q_{0}\right)+\frac{1}{2}\hat{\tilde{q}}\tilde{q}-y\hat{Q}Q-\hat{\tilde{Q}}\tilde{Q}-y\hat{M}\overline{W}-\hat{\tilde{M}}\overline{\tilde{W}}-y\left(\hat{S}S-\hat{\tilde{S}}\tilde{S}\right)+\right.\right.\label{eq:final_expr_RS}\\
 &  & \left.\left.-y\hat{D}\left(\frac{1}{2}Q+\frac{1}{2}\tilde{Q}-S-2D\right)\right)+\mathcal{G}_{S}+\alpha\mathcal{G}_{E}\right)\nonumber \\
\mathcal{G}_{S} & = & \int\mathcal{D}\tilde{z}\int\mathcal{D}z_{0}\log\left\{ \sum_{\tilde{l}}\exp\left(\left(\hat{\tilde{Q}}-\frac{1}{2}\hat{\tilde{q}}\right)\tilde{l}^{2}+\left(\tilde{z}\sqrt{\hat{\tilde{q}}-\frac{\hat{\tilde{S}}^{2}}{\hat{q}_{0}}}+z_{0}\frac{\hat{\tilde{S}}}{\sqrt{\hat{q}_{0}}}+\hat{\tilde{M}}\right)\tilde{l}\right)\times\right.\nonumber \\
 &  & \quad\left.\times\int\mathcal{D}z_{1}\left\{ \sum_{l}\exp\left(\left(\hat{Q}-\frac{1}{2}\hat{q}_{1}\right)l^{2}+\left(z_{0}\sqrt{\hat{q}_{0}}+z_{1}\sqrt{\hat{q}_{1}-\hat{q}_{0}}+\hat{M}+\left(\hat{S}-\hat{\tilde{S}}\right)\tilde{l}\right)l\right)\right\} ^{y}\right\} \nonumber \\
\mathcal{G}_{E} & = & \int\mathcal{D}z_{0}\int\mathcal{D}\tilde{z}\,\left\langle \log\left(\int\mathcal{D}z_{1}H\left(\frac{K-\overline{\xi}M-z_{0}\sqrt{\sigma_{\xi}^{2}q_{0}}-z_{1}\sqrt{\sigma_{\xi}^{2}\left(q_{1}-q_{0}\right)}}{\sqrt{\sigma_{\xi}^{2}\left(Q-q_{1}\right)}}\right)^{y}H\left(\tilde{C}\left(s,z_{0},z_{1},\tilde{z}\right)\right)\right)\right\rangle _{s}\nonumber 
\end{eqnarray}

\selectlanguage{english}%
where the stationarity condition implies the following saddle point
equations:

\begin{eqnarray}
 & \tilde{q}=-2\frac{\partial}{\partial\hat{\tilde{q}}}\mathcal{G}_{S};\quad\tilde{Q}=\frac{\partial}{\partial\hat{\tilde{Q}}}\mathcal{G}_{S};\quad q_{0}=-\frac{2}{y^{2}}\frac{\partial}{\partial\hat{q}_{0}}\mathcal{G}_{S};\quad q_{1}=\frac{2}{y\left(y-1\right)}\frac{\partial}{\partial\hat{q_{1}}}\mathcal{G}_{S};\quad Q=\frac{1}{y}\frac{\partial}{\partial\hat{Q}}\mathcal{G}_{S};\label{eq:sys2}\\
 & \tilde{S}=-\frac{1}{y}\frac{\partial}{\partial\hat{\tilde{S}}}\mathcal{G}_{S};\quad S=\frac{Q}{2}+\frac{\tilde{Q}}{2}-2D;\quad\overline{W}=\frac{1}{y}\frac{\partial}{\partial\hat{M}}\mathcal{G}_{S};\quad\overline{\tilde{W}}=\frac{\partial}{\partial\hat{\tilde{M}}}\mathcal{G}_{S};\quad0=\frac{1}{y}\frac{\partial}{\partial\hat{S}}\mathcal{G}_{S}-S;\nonumber \\
 & \hat{\tilde{q}}=-2\alpha\frac{\partial}{\partial\tilde{q}}\mathcal{G}_{E};\quad\hat{\tilde{Q}}=-y\frac{\hat{D}}{2}+\alpha\frac{\partial}{\partial\tilde{Q}}\mathcal{G}_{E};\quad\hat{q}_{0}=-\frac{2\alpha}{y^{2}}\frac{\partial}{\partial q_{0}}\mathcal{G}_{E};\quad\hat{q}_{1}=\frac{2\alpha}{y\left(y-1\right)}\frac{\partial}{\partial q_{1}}\mathcal{G}_{E};\quad\hat{Q}=-\frac{\hat{D}}{2}+\frac{\alpha}{y}\frac{\partial}{\partial Q}\mathcal{G}_{E};\nonumber \\
 & \quad\hat{D}=\hat{S}-\frac{\alpha}{y}\frac{\partial}{\partial S}\mathcal{G}_{E};\quad\hat{\tilde{S}}=-\frac{\alpha}{y}\frac{\partial}{\partial\tilde{S}}\mathcal{G}_{E};\quad\hat{M}=0;\quad0=\frac{\partial}{\partial M}\mathcal{G}_{E};\quad0=\frac{\partial}{\partial\tilde{M}}\mathcal{G}_{E}.\nonumber 
\end{eqnarray}

We are thus left with a system of $19$ coupled equations and three
control parameters $\alpha$, $y$ and $D$. 

For each couple of $\alpha$ and $D$, the sought value of the inverse
temperature $y^{\star}$ corresponding to a vanishing external entropy
$\Sigma_{RC}$ (eq.~\eqref{eq:ext_entropy}) can be found by interpolating
between different saddle point solutions at varying values of $y$.
As in the case for the Franz-Parisi potential the saddle point equations
are best controlled by fixing the conjugate parameter $\hat{Q}$ and
consequently determining the correspondent value of $D$. In this
way the number of implicit equations to be solved via Newton's method
is minimized. The saddle point solutions can then be found by iterating
the equations~\eqref{eq:sys2}.

\section{RS solution, large $y$ limit\label{sec:RS-solution-large-y}}

We study the final RS expression for the large-deviation free energy
density in the $y\to\infty$ limit, where further simplifications
can be made. We have seen that when $\alpha$ is sufficiently high
and $D$ approaches zero this analysis returns some unphysical results
that need to be corrected by introducing a different Ansatz for the
order parameters. Still, for a large range of values for $\alpha$
this limit is in good agreement with the more involved analyses and
can provide some insight into the physical phenomena under study.

The first major simplification comes from the observation that the
$\mathbb{X}_{\xi,\sigma}\left(\tilde{W},K\right)$ constraint on the
reference configuration effectively disappears when the temperature
$y$ becomes large: the saddle point solution for the order parameters
describing the clustered solutions remains unaltered when this constraint
is completely removed. The reason is the following: in the expression
for $\mathcal{G}_{E}$ of eq.~\eqref{eq:final_expr_RS}, the expressions
$H\left(\tilde{C}\left(s,z_{0},z_{1},\tilde{z}\right)\right)$ are
not elevated to the power of $y$, and thus they become effectively
irrelevant, implying that $\mathcal{G}_{E}$ is constant with respect
to the order parameters $\tilde{Q}$, $\tilde{q}$, $\tilde{S}$ and
$\tilde{M}$. In turn, looking at the saddle point equations, this
means that $\hat{\tilde{q}}$, $\hat{\tilde{S}}$, $\hat{\tilde{M}}$
are all $0$, and that $\hat{\tilde{Q}}=-y\frac{\hat{D}}{2}$. Furthermore,
the term $\hat{\tilde{M}}\overline{\tilde{W}}$ is also negligible,
since it is not scaled with $y$.

Thus the large $y$ case formally be obtained from the final expression~\eqref{eq:final_expr_RS}
by setting to zero the order parameters describing the planted configuration
$\tilde{q}$, $\tilde{S}$, $\tilde{M}$, $\overline{\tilde{W}}$
and their conjugates (even though the order parameters are not $0$,
and their value can be obtained by carefully performing the limit).
The only surviving term is the $L^{2}$-norm $\tilde{Q}$, which is
also involved in the distance constraint. Moreover the integration
over $\tilde{z}$ in the $\mathcal{G}_{S}$ and $\mathcal{G}_{E}$
terms can be carried out analytically. The final expression for the
free entropy in this limit is thus the same as for the unconstrained
case, namely:

\selectlanguage{british}%
\begin{eqnarray}
\Phi_{RU}\left(D,y\right) & = & -\left(\left(\frac{y}{2}\hat{q}_{1}q_{1}-\frac{y^{2}}{2}\left(\hat{q}_{1}q_{1}-\hat{q}_{0}q_{0}\right)+y\hat{Q}Q-\hat{\tilde{Q}}\tilde{Q}-y\hat{M}\overline{W}-y\hat{S}S+\right.\right.\\
 &  & \left.\left.-y\hat{D}\left(\frac{1}{2}Q+\frac{1}{2}\tilde{Q}-S-2D\right)\right)+\mathcal{G}_{S}+\alpha\mathcal{G}_{E}\right)\nonumber \\
\mathcal{G}_{S} & = & \int\mathcal{D}z_{0}\log\left\{ \sum_{\tilde{l}}e^{\hat{\tilde{Q}}\tilde{l}^{2}}\int\mathcal{D}z_{1}\left\{ \sum_{l}\exp\left(\left(\hat{Q}-\frac{1}{2}\hat{q}_{1}\right)l^{2}+\left(z_{0}\sqrt{\hat{q}_{0}}+z_{1}\sqrt{\hat{q}_{1}-\hat{q}_{0}}+\hat{M}+\hat{S}\tilde{l}\right)l\right)\right\} ^{y}\right\} \nonumber \\
\mathcal{G}_{E} & = & \int\mathcal{D}z_{0}\left\langle \log\left(\int\mathcal{D}z_{1}H\left(\frac{K-s\overline{G}M-z_{0}\sqrt{\sigma_{G}^{2}q_{0}}-z_{1}\sqrt{\sigma_{G}^{2}\left(q_{1}-q_{0}\right)}}{\sqrt{\sigma_{G}^{2}\left(Q-q_{1}\right)}}\right)^{y}\right)\right\rangle _{s}\nonumber 
\end{eqnarray}

\selectlanguage{english}%
In order to take the $y\to\infty$ limit we need to make a self-consistent
Ansatz for the scaling of some order parameters with $y$: the difference
between the two overlaps $\left(q_{1}-q_{0}\right)$ vanishes in this
limit, so we can define $q_{0}=q$ and consider the scaling $q_{1}\to q+\frac{\delta q}{y}$.
Similarly we can pose $\hat{q}_{0}=\hat{q}$, $\hat{q}_{1}\to\hat{q}+\frac{\delta\hat{q}}{y}$.
Note that $q_{1}\to q_{0}$ also implies $\tilde{q}\to\tilde{Q}$,
which could be verified by a first-order expansion in $y^{-1}$.

We can now evaluate the $\int\mathcal{D}z_{1}$ integrals appearing
in the energetic and the entropic terms, resorting to a first order
saddle point approximation: for this purpose we need to rescale the
integration variable $z_{1}^{\prime}=\sqrt{y}z_{1}$. 

The $z_{1}$ integrals and the summation over $\tilde{l}$ in the
entropic term are thus replaced by maximum functions, obtaining:

\begin{eqnarray}
 &  & \lim_{y\to\infty}\Phi_{RU}\left(D,y\right)\approx\\
 &  & \approx\lim_{y\to\infty}y\left\{ \frac{1}{2}\hat{q}q-\frac{1}{2}\delta\hat{q}q-\frac{1}{2}\hat{q}\delta q-\hat{Q}Q-\frac{\hat{\tilde{Q}}\tilde{Q}}{y}-\hat{M}\overline{W}-\hat{S}S-\frac{1}{2}\hat{D}Q-\frac{1}{2}D\tilde{Q}+\hat{D}S+2\hat{D}D+\right.\nonumber \\
 &  & \quad+\int\mathcal{D}z_{0}\max_{\tilde{l}}\left(\frac{\hat{\tilde{Q}}}{y}\tilde{l}^{2}+\max_{z_{1}}\left(-\frac{z_{1}^{2}}{2}+\log\left(\sum_{l}\exp\left(\left(\hat{Q}-\frac{1}{2}\hat{q}\right)l^{2}+\left(z_{0}\sqrt{\hat{q}}+z_{1}\sqrt{\delta\hat{q}}+\hat{M}+S\tilde{l}\right)l\right)\right)\right)\right)+\nonumber \\
 &  & \quad\left.+\alpha\int\mathcal{D}z_{0}\left\langle \max_{z_{1}}\left(-\frac{z_{1}^{2}}{2}+\log\left(H\left(\frac{K-s\overline{\xi}M-z_{0}\sqrt{\sigma_{\xi}^{2}q}-z_{1}\sqrt{\sigma_{\xi}^{2}\delta q}}{\sqrt{\sigma_{\xi}^{2}\left(Q-q\right)}}\right)\right)\right)\right\rangle _{s}\right\} \nonumber 
\end{eqnarray}
where the constant vanishing term $\frac{\log y}{y}$ was neglected.
We kept the subleading term $\frac{\hat{\tilde{Q}}\tilde{Q}}{y}$
so we could derive the trivial saddle point equation:

\begin{eqnarray}
\hat{\tilde{Q}} & = & -\frac{1}{2}y\hat{D}
\end{eqnarray}

Since also the equation $\hat{S}=\hat{D}$ holds, after the appropriate
substitutions both $\tilde{Q}$ and $S$ come out from the picture
and can be ignored.

With the definitions:

\begin{eqnarray}
 & \mathcal{W}_{l}\left(z_{1},\tilde{l}\right)=\exp\left(\left(\hat{Q}-\frac{1}{2}\hat{q}\right)l^{2}+\left(z_{0}\sqrt{\hat{q}}+z_{1}\sqrt{\delta\hat{q}}+\hat{M}+S\tilde{l}\right)l\right)\\
 & \text{argH}\left(s,z_{1}\right)=\frac{K-s\overline{\xi}M-z_{0}\sqrt{\sigma_{\xi}^{2}q}-z_{1}\sqrt{\sigma_{\xi}^{2}\delta q}}{\sqrt{\sigma_{\xi}^{2}\left(Q-q\right)}}
\end{eqnarray}
the other saddle point equations turn out to be:

\begin{eqnarray}
 & 0=D-\frac{Q}{4}+\frac{1}{2}\int\mathcal{D}z_{0}\left(-\frac{\left(\tilde{l}^{\star}\right)^{2}}{2}+\frac{\tilde{l}^{\star}z_{1}^{\star}}{\sqrt{\delta\hat{q}}}\right);\quad\hat{M}=0;\\
 & q=2\int\mathcal{D}z_{0}\frac{\left(z_{1}^{\star}\right)^{2}}{2\delta\hat{q}};\quad Q=\int\mathcal{D}z_{0}\frac{\sum_{l}\mathcal{W}_{l}\left(z_{1}^{\star},\tilde{l}^{\star}\right)l^{2}}{\sum_{l}\mathcal{W}_{l}\left(z_{1}^{\star},\tilde{l}^{\star}\right)};\\
 & \delta q=q+2\int\mathcal{D}z_{0}\frac{\sum_{l}\mathcal{W}_{l}\left(z_{1}^{\star},\tilde{l}^{\star}\right)\left(-\frac{l^{2}}{2}+\frac{z_{0}l}{2\sqrt{\hat{q}}}\right)}{\sum_{l}\mathcal{W}_{l}\left(z_{1}^{\star},\tilde{l}^{\star}\right)};\\
 & \hat{q}=2\alpha\left(f^{\prime}\int\mathcal{D}z_{0}\left(z_{1}^{\star}\right)^{2}+\left(1-f^{\prime}\right)\int\mathcal{D}z_{0}\left(z_{1}^{\star}\right)^{2}\right);\\
 & 0=\int\mathcal{D}z_{0}\left\langle -\frac{s\overline{\xi}z_{1}^{\star}}{\sqrt{\sigma_{\xi}^{2}\delta q}}\right\rangle _{s};\quad\overline{W}=\int\mathcal{D}z_{0}\frac{z_{1}^{\star}}{\sqrt{\delta\hat{q}}};\\
 & \delta\hat{q}=\hat{q}+2\alpha\int\mathcal{D}z_{0}\left\langle z_{1}^{\star}\left(\frac{\text{argH}\left(s,z_{1}^{\star}\right)}{\sqrt{\sigma_{\xi}^{2}(Q-q)}}+\frac{z_{0}}{\sqrt{q}}\right)\right\rangle _{s};\\
 & \hat{D}=-2\hat{Q}+2\alpha\int\mathcal{D}z_{0}\left\langle z_{1}^{\star}\text{argH}\left(s,z_{1}^{\star}\right)\right\rangle _{s};
\end{eqnarray}

where $z_{1}^{\star}$ and $\tilde{l}^{\star}$ are to be intended
as functions of $z_{0}$, corresponding to the values of $z_{1}$
and $\tilde{l}$ that maximize $\mathcal{G}_{S}$ or $\mathcal{G}_{E}$
at that fixed $z_{0}$. 

This yields a system of $9$ coupled equations with two control parameters
$\alpha$ and $D$, which can again be solved by iteration, resorting
to Newton's method for the implicit equations. The parameter $\hat{Q}$
is still the best practical choice for the control parameter, being
a bijective function of $D$ and allowing for a reduction of the number
of Newton's routines at each iteration.

\section{External 1RSB Ansatz, unconstrained case, large y limit\label{sec:AppendixD-1RSB}}

Starting from expression \eqref{eq:before_ansatz_RS} for the replicated
volume in the constrained reweighted measure, as an alternative we
can opt for a 1RSB Ansatz for the planted configurations. From a geometrical
point of view this scheme describes a situation where the $n$ replicas
are organized in $\frac{n}{m}$ blocks of $m$ replicas each, $m$
being the Parisi 1RSB parameter over which we will subsequently optimize.
This leads to an expression which is formally similar to a 2RSB description,
analogously to how the RS case is formally similar to a 1RSB description.

We therefore need to introduce the multi-index $c=\left(\alpha,\beta\right)$,
where $\alpha\in\left\{ 1,...,n/m\right\} $ labels a block of $m$
replicas, and $\beta\in\left\{ 1,...,m\right\} $ indexes the replicas
inside the block. This induces a slightly more complicated structure
for the overlap matrix $q^{ca,db}$:
\begin{equation}
q^{\alpha\beta,a;\alpha^{\prime}\beta^{\prime},b}=\begin{cases}
Q & \textrm{if}\,\alpha=\alpha^{\prime},\beta=\beta^{\prime},a=b\\
q_{2} & \textrm{if}\,\alpha=\alpha^{\prime},\beta=\beta^{\prime},a\ne b\\
q_{1} & \textrm{if}\,\alpha=\alpha^{\prime},\beta\ne\beta^{\prime}\\
q_{0} & \textrm{if}\,\alpha\ne\alpha^{\prime}
\end{cases}\label{eq:1rsb_ansatz}
\end{equation}
and similarly for the conjugated parameter matrix $\hat{q}^{ca,db}$.
In this case, we drop the constraint $\mathbb{X}_{\xi,\sigma}\left(\tilde{W},K\right)$
on the reference configurations, thus getting rid of the parameters
$\tilde{q}$, $\tilde{S}$, $\overline{\tilde{W}}$, $\tilde{M}$
and their conjugates. The Ansatz for the remaining order parameters
remains unchanged from the RS case. 

Following step by step the calculations presented in the Appendix
of \cite{baldassi_local_2016} one can obtain the following expression
for the free entropy density $\Phi_{RU}\left(D,y\right)$:

\begin{eqnarray}
\Phi_{RU}\left(D,y\right) & \approx & -\left(y^{2}\frac{m}{2}\hat{q}_{0}q_{0}-y^{2}\frac{m-1}{2}\hat{q}_{1}q_{1}-y\frac{y-1}{2}\hat{q}_{2}q_{2}-y\hat{Q}Q-y\hat{D}\left(\frac{1}{2}Q-2D\right)+\mathcal{G}_{S}+\alpha\mathcal{G}_{E}\right)\\
\mathcal{G}_{S} & = & \frac{1}{m}\int\mathcal{D}z_{0}\log\int\mathcal{D}z_{1}Z\left(z_{0},z_{1}\right)^{m}\nonumber \\
Z\left(z_{0},z_{1}\right) & = & \int\mathcal{D}z_{2}\ \sum_{\tilde{l}}e^{-\frac{1}{2}y\hat{D}\tilde{l}^{2}}\bigg[\sum_{l}\exp\left(\left(\hat{Q}-\frac{1}{2}\hat{q}_{2}\right)l^{2}+\left(z_{0}\sqrt{\hat{q}_{0}}+z_{1}\sqrt{\hat{q}_{1}-\hat{q}_{0}}+z_{2}\sqrt{\hat{q}_{2}-\hat{q}_{1}}+\hat{M}+\hat{D}\,\tilde{l}\right)l\right)\bigg]^{y}\nonumber \\
\mathcal{G}_{E} & = & \frac{1}{m}\int\negthickspace\mathcal{D}z_{0}\,\left\langle \log\int\negthickspace\mathcal{D}z_{1}\left[\int\mathcal{D}z_{2}\ H\left(\frac{K-s\overline{\xi}M-z_{0}\sqrt{\sigma_{\xi}^{2}q_{0}}-z_{1}\sqrt{\sigma_{\xi}^{2}\left(q_{1}-q_{0}\right)}-z_{2}\sqrt{\sigma_{\xi}^{2}\left(q_{2}-q_{1}\right)}}{\sqrt{\sigma_{\xi}^{2}\left(Q-q_{2}\right)}}\right)^{y}\right]^{m}\right\rangle _{s}\nonumber 
\end{eqnarray}
where we already substituted the trivial saddle point equations:

\begin{equation}
\hat{\tilde{Q}}=-\frac{1}{2}y\hat{D};\quad\hat{S}=\hat{D};\quad\hat{M}=0
\end{equation}

When we send $y\to\infty$ this time we must pose the scalings $m\to\frac{x}{y}$,
$q_{2}\to q_{1}+\frac{\delta q}{y}$ and $\hat{q}_{2}\to\hat{q}_{1}+\frac{\delta\hat{q}}{y}$;
using again a saddle point approximation for the $\int\mathcal{D}z_{2}$
integral, to the leading order in $y$ we find:

\begin{eqnarray}
\lim_{y\to\infty}\Phi_{RU}\left(D,y\right) & \approx & \lim_{y\to\infty}-y\left(\frac{x}{2}\left(\hat{q}_{0}q_{0}-\hat{q}_{1}q_{1}\right)-\frac{1}{2}\left(\delta\hat{q}\,q_{1}+\hat{q}_{1}\,\delta q\right)+\frac{1}{2}\hat{q}_{1}q_{1}-\hat{Q}Q+\right.\\
 &  & \qquad\qquad\left.-\hat{D}\left(\frac{1}{2}Q-2D\right)+\mathcal{G}_{S}^{\infty}+\alpha\mathcal{G}_{E}^{\infty}\right)\nonumber \\
\mathcal{G}_{S}^{\infty} & = & \frac{1}{x}\int\mathcal{D}z_{0}\ \log\int\mathcal{D}z_{1}\ e^{xA_{S}\left(z_{0},z_{1}\right)}\nonumber \\
A_{S}\left(z_{0},z_{1}\right) & = & \max_{\tilde{l},z_{2}}\bigg\{-\frac{\hat{D}\tilde{l}^{2}}{2}-\frac{z_{2}^{2}}{2}+\log\sum_{l}e^{\left(\hat{Q}-\frac{1}{2}\hat{q}_{1}\right)l^{2}+\left(z_{0}\sqrt{\hat{q}_{0}}+z_{1}\sqrt{\hat{q}_{1}-\hat{q}_{0}}+z_{2}\sqrt{\delta\hat{q}}+\hat{D}\,\tilde{l}\right)l}\bigg\}\nonumber \\
\mathcal{G}_{E}^{\infty} & = & \frac{1}{x}\int\mathcal{D}z_{0}\left\langle \log\int\mathcal{D}z_{1}\ e^{xA_{E}\left(s,z_{0},z_{1}\right)}\right\rangle _{s}\nonumber \\
A_{E}\left(s,z_{0},z_{1}\right) & = & \max_{z_{2}}\bigg\{-\frac{z_{2}^{2}}{2}+\log H\left(\frac{K-s\overline{\xi}M-z_{0}\sqrt{\sigma_{\xi}^{2}q_{0}}-z_{1}\sqrt{\sigma_{\xi}^{2}\left(q_{1}-q_{0}\right)}-z_{2}\sqrt{\sigma_{\xi}^{2}\,\delta q}}{\sqrt{\sigma_{\xi}^{2}\left(Q-q_{1}\right)}}\right)\bigg\}\nonumber 
\end{eqnarray}

where the order parameters take the value obtained by solving the
saddle point equations:

\begin{eqnarray}
 & \hat{M}=0;\quad\hat{q}_{0}=-\frac{2\alpha}{x}\frac{\partial\mathcal{G}_{E}}{\partial q_{0}};\quad\hat{q}_{1}=2\alpha\frac{\partial\mathcal{G}_{E}}{\partial\delta q};\quad\frac{\partial\mathcal{G}_{E}}{\partial M}=0;\label{eq:sys3}\\
 & \delta\hat{q}=2\alpha\frac{\partial\mathcal{G}_{E}}{\partial q_{1}}+(1-x)\hat{q}_{1};\quad\hat{D}=-2\hat{Q}+2\alpha\frac{\partial\mathcal{G}_{E}}{\partial Q};\quad q_{1}=2\frac{\partial\mathcal{G}_{S}}{\partial\delta q};\nonumber \\
 & \delta q=2\frac{\partial\mathcal{G}_{S}}{\partial\hat{q}_{1}}+(1-x)q_{1};\quad Q=\frac{\partial\mathcal{G}_{S}}{\partial\hat{Q}};\quad q_{0}=-\frac{2}{x}\frac{\partial\mathcal{G}_{S}}{\partial\hat{q}_{0}};\quad0=D-\frac{Q}{4}-\frac{1}{2}\frac{\partial\mathcal{G}_{S}}{\partial\hat{D}}\nonumber 
\end{eqnarray}

Differently from the previous case, in addition to $\alpha$ and $D$
in this system of equations we have the control parameter $x$, which
we can optimize on by requiring the saddle point condition $\frac{\partial\Phi_{RU}}{\partial x}=0$.
Since the other saddle point equations are sensitive even to small
changes of its value, a good way of finding this solution is to reach
convergence of the other order parameters by iterating the equations
in~\eqref{eq:sys3} at fixed values of $x$, and then interpolate
to find the zero of the function $\frac{\partial\Phi_{RU}}{\partial x}$$\left(x\right)$.
Again, instead of spanning the values of $D$ directly, it is better
to use $\hat{Q}$ as a control parameter.

\end{document}